%% file: main.tex
\input{LaTeX/config}
\begin{document}

\input{LaTeX/chapters/chapter_0_ABSTRACT}




\maketitle
\renewcommand{\shortauthors}{Ranxu Zhang et al.}

\input{LaTeX/chapters/chapter_1_INTRODUCTION}
\input{LaTeX/chapters/chapter_2_RELATED_WORK}
\input{LaTeX/chapters/chapter_3_PRELIMINARY}
\input{LaTeX/chapters/chapter_4_METHOD}
\input{LaTeX/chapters/chapter_5_EXPERIMENTS}
\input{LaTeX/chapters/chapter_6_CONCLUSION}

\begin{acks}
This work was supported in part by the Natural Science Foundation of Anhui Province (Grant No. 2508085QF211), the National Natural Science Foundation of China (Grant No. 62506348), New Generation Artificial Intelligence-National Science and Technology Major Project (Grant No. 2025ZD0122601), the Opening
Foundation of State Key Laboratory of Cognitive Intelligence, iFLYTEK (COGOS-2025HE02), the CCF-1688 Yuanbao Cooperation Fund (Grant No. CCF-Alibaba2025005).
\end{acks}
\bibliographystyle{ACM-Reference-Format}
\balance
\bibliography{ref}

\input{LaTeX/chapters/chapter_7_APPENDIX}

\end{document}

%% file: LaTeX/config.tex
\documentclass[sigconf]{acmart}
\usepackage{graphicx}
\usepackage{multirow}
\usepackage{makecell}
\usepackage{balance}
\AtBeginDocument{%
  }

\copyrightyear{2026}
\acmYear{2026}
\setcopyright{cc}
\setcctype{by}
\acmConference[WWW '26]{Proceedings of the ACM Web Conference 2026}{April 13--17, 2026}{Dubai, United Arab Emirates}
\acmBooktitle{Proceedings of the ACM Web Conference 2026 (WWW '26), April 13--17, 2026, Dubai, United Arab Emirates}
\acmPrice{}
\acmDOI{10.1145/3774904.3792495}
\acmISBN{979-8-4007-2307-0/2026/04}

\title{MCLMR: A Model-Agnostic Causal Learning Framework for Multi-Behavior Recommendation}

\author{Ranxu Zhang}
\orcid{0009-0006-7222-2149}
\affiliation{
  \institution{School of Artificial Intelligence and Data Science, University of Science and Technology of China}
  \city{Hefei}
  \country{China}
}
\email{zkd_zrx@mail.ustc.edu.cn}

\author{Junjie Meng}
\orcid{0009-0006-5709-9963}
\affiliation{
  \institution{School of Artificial Intelligence and Data Science, University of Science and Technology of China}
  \city{Hefei}
  \country{China}
}
\email{mengjre@gmail.com}

\author{Ying Sun}
\orcid{0000-0002-4763-6060}
\affiliation{
  \institution{Artificial Intelligence Thrust, The Hong Kong University of Science and Technology (Guangzhou)}
  \city{Guangzhou}
  \country{China}
}
\email{sunyinggilly@gmail.com}

\author{Ziqi Xu}
\orcid{0000-0003-1748-5801}
\affiliation{%
  \institution{RMIT University}
  \city{Melbourne}
  \country{Australia}
}
\email{ziqi.xu@rmit.edu.au}

\author{Bing Yin}
\orcid{0000-0002-4042-7522}
\affiliation{%
  \department{State Key Laboratory of Cognitive Intelligence}
  \institution{iFLYTEK Research, iFLYTEK}
  \city{Hefei}
  \country{China}
}
\email{bingyin@iflytek.com}

\author{Hao Li}
\orcid{0009-0002-2556-4841}
\affiliation{%
  \institution{iFLYTEK Research, iFLYTEK}
  \city{Hefei}
  \country{China}
}
\email{haoli5@iflytek.com}

\author{Yanyong Zhang}
\orcid{0000-0001-9046-798X}
\affiliation{
  \institution{School of Artificial Intelligence and Data Science, University of Science and Technology of China}
  \city{Hefei}
  \country{China}
}
\email{yanyongz@ustc.edu.cn}

\author{Chao Wang}
\authornote{Chao Wang is the corresponding author.}
\orcid{0000-0001-7717-447X}
\affiliation{
  \institution{School of Artificial Intelligence and Data Science, University of Science and Technology of China}
  \department{State Key Laboratory of Cognitive Intelligence}
  \city{Hefei}
  \country{China}
}
\email{wangchaoai@ustc.edu.cn}

%% file: LaTeX/chapters/chapter_0_ABSTRACT.tex
\begin{abstract}
Multi-Behavior Recommendation (MBR) leverages multiple user interaction types (e.g., views, clicks, purchases) to enrich preference modeling and alleviate data sparsity issues in traditional single-behavior approaches. However, existing MBR methods face fundamental challenges: they lack principled frameworks to model complex confounding effects from user behavioral habits and item multi-behavior distributions, struggle with effective aggregation of heterogeneous auxiliary behaviors, and fail to align behavioral representations across semantic gaps while accounting for bias distortions. To address these limitations, we propose MCLMR, a novel model-agnostic causal learning framework that can be seamlessly integrated into various MBR architectures. MCLMR first constructs a causal graph to model confounding effects and performs interventions for unbiased preference estimation. Under this causal framework, it employs an Adaptive Aggregation module based on Mixture-of-Experts to dynamically fuse auxiliary behavior information and a Bias-aware Contrastive Learning module to align cross-behavior representations in a bias-aware manner. Extensive experiments on three real-world datasets demonstrate that MCLMR achieves significant performance improvements across various baseline models, validating its effectiveness and generality. All data and code will be made publicly available. For anonymous review, our code is available at the following the link: https://github.com/gitrxh/MCLMR.

\end{abstract}

\ccsdesc[500]{Information systems~Recommender systems}
\ccsdesc[300]{Computing methodologies~Causal reasoning and diagnostics}

%% file: LaTeX/chapters/chapter_1_INTRODUCTION.tex
\section{Introduction}

Recommender systems~\cite{wang2021personalized,huang2021recent,lightgcn,wang2025face} are essential for information discovery. While Traditional Collaborative Filtering (CF)~\cite{cfsurvey,ncf,wang2018confidence,wang2020setrank} effectively models preferences, it typically relies on single-behavior data, causing data sparsity. Multi-Behavior Recommendation (MBR)~\cite{mbgcn,matn,nmtr} addresses this by leveraging diverse interactions (e.g., views, clicks) to enrich preference modeling. However, existing MBR approaches face a fundamental challenge: multi-behavior data inherently contains multi-behavior biases, where different users exhibit distinct behavioral patterns and different items receive varying behavior type distributions. Ignoring such bias modeling leads to suboptimal preference learning and compromised recommendation quality.

To leverage multiple behaviors, researchers have proposed various MBR approaches. Early methods like CMF~\cite{cmf} employ matrix factorization to learn shared representations across behaviors. Recently, Graph Neural Network (GNN)-based models have become mainstream, with MBGCN~\cite{mbgcn} propagating information through behavior-specific layers and CRGCN~\cite{crgcn} designing cascaded networks to model behavioral dependencies. However, these methods primarily focus on capturing behavioral correlations while overlooking multi-behavior biases. Meanwhile, traditional debiasing methods address single-behavior scenarios through techniques like Inverse Propensity Scoring (IPS) or causal inference to mitigate popularity bias~\cite{10.1145/3564284} and exposure bias~\cite{2020arXiv200615772A, 10.1145/3627673.3679763}. Recent efforts have extended causal reasoning to multi-behavior settings: CVID~\cite{10.1145/3745023} employs diffusion models to infer latent confounders, while CMSR~\cite{cmsr} constructs causal graphs to model behavioral dependencies. Nevertheless, these approaches either lack explicit structural modeling of inter-behavior causality or insufficiently capture the dual confounding effects from both user behavioral habits and item multi-behavior distributions. How to systematically disentangle these multi-faceted biases remains an open challenge.


Hence, we identify three key challenges for multi-behavior debiasing. First, complex confounding behavioral interaction tendencies exist where popularity intertwines with behavior-specific biases (e.g., popular items attract more views regardless of preference). Second, heterogeneous behavioral patterns require dynamic aggregation, as auxiliary behaviors vary in informativeness for specific user-item pairs. Third, significant semantic gaps and bias distortions hinder effective cross-behavior representation alignment. Effectively aligning these behavioral representations in a bias-aware manner while preserving distinct semantic meanings is crucial.

Based on the above considerations, we propose a novel, model-agnostic causal learning framework for multi-behavior recommendation, named \textbf{MCLMR}, which is designed as a universal plug-in to enhance existing MBRs models. Specifically, we first construct a causal graph to model the complex confounding effects in multi-behavior data and derive unbiased preference estimation through causal intervention. Under this causal framework, MCLMR addresses the remaining two challenges through: (1) an \textbf{Adaptive Aggregation module} based on a Mixture-of-Experts (MoE) architecture, which dynamically learns to fuse information from heterogeneous auxiliary behaviors in a user-item-specific manner to maximize positive transfer and suppress negative transfer; and (2) a \textbf{Bias-aware Contrastive Learning module} that employs a novel user-item dual alignment strategy to align preference representations across behavioral spaces while accounting for the distortion caused by biases. Our main contributions are as follows:
\begin{itemize}
    \item We propose \textbf{MCLMR}, a novel model-agnostic causal framework that can be seamlessly integrated into various MBRs architectures to enhance their performance.
    \item We construct a causal graph to model confounding effects in multi-behavior data and derive principled intervention strategies for unbiased preference estimation.
    \item We design an Adaptive Aggregation module based on MoE architecture to dynamically fuse auxiliary behavior information and a Bias-aware Contrastive Learning module to align cross-behavior representations.
    \item Extensive experiments on three public datasets validate the superiority of our framework and the effectiveness of each core component.
\end{itemize}

%% file: LaTeX/chapters/chapter_2_RELATED_WORK.tex
\section{Related Work}

\subsection{Multi-behavior Recommendation}

Multi-behavior Recommendation (MBR)~\cite{WanyanHM0C25,kmclr,Ma0RHC25} integrates diverse interactions to alleviate target behavior sparsity. Early methods like CMF~\cite{cmf} jointly factorized behavior matrices via shared latent factors. Recently, GNN-based models~\cite{wang2021variable} have become dominant. MBGCN~\cite{mbgcn} learns behavior strength by propagating information on a unified heterogeneous graph, while CRGCN~\cite{crgcn} employs a cascaded GCN structure to model sequential dependencies and refine preferences. However, MBGCN often overlooks the semantic gap between behaviors, and CRGCN models raw interactions directly. Most existing methods fail to explicitly address the inherent behavioral biases that confound true user intent.

\subsection{Debiased Recommendation}

Debiased recommendation~\cite{ZhangYCLLXZ25,chen2021biasdebiasrecommendersystem, luo2024ci4rs,liu2025rethinking} addresses biases caused by feedback loops. Inverse Propensity Scoring (IPS)~\cite{CounterfactualReasoningBottou2013} is a common technique; Multi-IPW and Multi-DR~\cite{DBLP:journals/corr/abs-1910-09337} extend IPS to multi-behavior scenarios, though they often require normalization (SNIPS) to mitigate high variance. Causal inference~\cite{XuCLLLW23,2018arXiv180806581W, krauth2022breakingfeedbackloopsrecommender,DBLP:conf/iclr/XuCLL0Y24} offers a deeper approach. Methods like PDA~\cite{10.1145/3404835.3462875}, CausE~\cite{10.1145/3240323.3240360}, and MACR~\cite{wei2021model} use causal intervention to handle popularity bias, while DCCL~\cite{10.1145/3543873.3584637} employs contrastive learning to disentangle interest from conformity. To extend causal reasoning~\cite{10.1016/j.ins.2024.120834} to MBR, CVID~\cite{10.1145/3745023} infers latent confounders via diffusion models~\cite{zhu2024graph}, and CMSR~\cite{cmsr} uses causal graphs to model behavioral dependencies. However, CVID lacks explicit structural modeling of inter-behavior causality, while CMSR insufficiently captures complex unobserved confounders. Effectively handling both within a unified framework remains an open challenge, which is the core issue this paper aims to address.

%% file: LaTeX/chapters/chapter_3_PRELIMINARY.tex
\section{Preliminaries}

\paragraph{Basic Notations}
In our framework, we denote the set of users as $U = \{u_1, u_2, \dots, u_M\}$ and the set of items as $V = \{v_1, v_2, \dots, v_N\}$, where $M$ and $N$ are the total number of users and items, respectively.

\paragraph{Multi-Behavior Interaction Data}
We consider $K$ behavior types with inherent sequential dependencies (e.g., from 'view' to 'cart', then 'purchase'). The user interaction data forms a set of interaction subgraphs $\mathcal{G} = \{\mathcal{G}_1, \dots, \mathcal{G}_K\}$, where each $\mathcal{G}_k$ contains all observed interactions for behavior type $k$.
We explicitly define the final behavior $K_t$ (e.g., 'purchase') as the \textbf{target behavior}, which is the ultimate prediction goal. The preceding behaviors $\{1, \dots, K_t-1\}$ serve as auxiliary behaviors to enhance the target prediction.

\paragraph{Model-Agnostic Setting and Embeddings}
Designed as a model-agnostic module, our framework can be integrated with any backbone multi-behavior model (e.g., CRGCN, BCIPM). It takes the backbone's learned embeddings as input. Specifically, we denote $\mathbf{e}_{u,k}, \mathbf{e}_{i,k} \in \mathbb{R}^d$ as the behavior-specific embeddings for user $u$ and item $i$ under behavior $k$. These embeddings capture user preferences and item attributes at different behavioral stages.

\begin{figure*}[t!]
    \centering
    \includegraphics[width=0.85\textwidth]{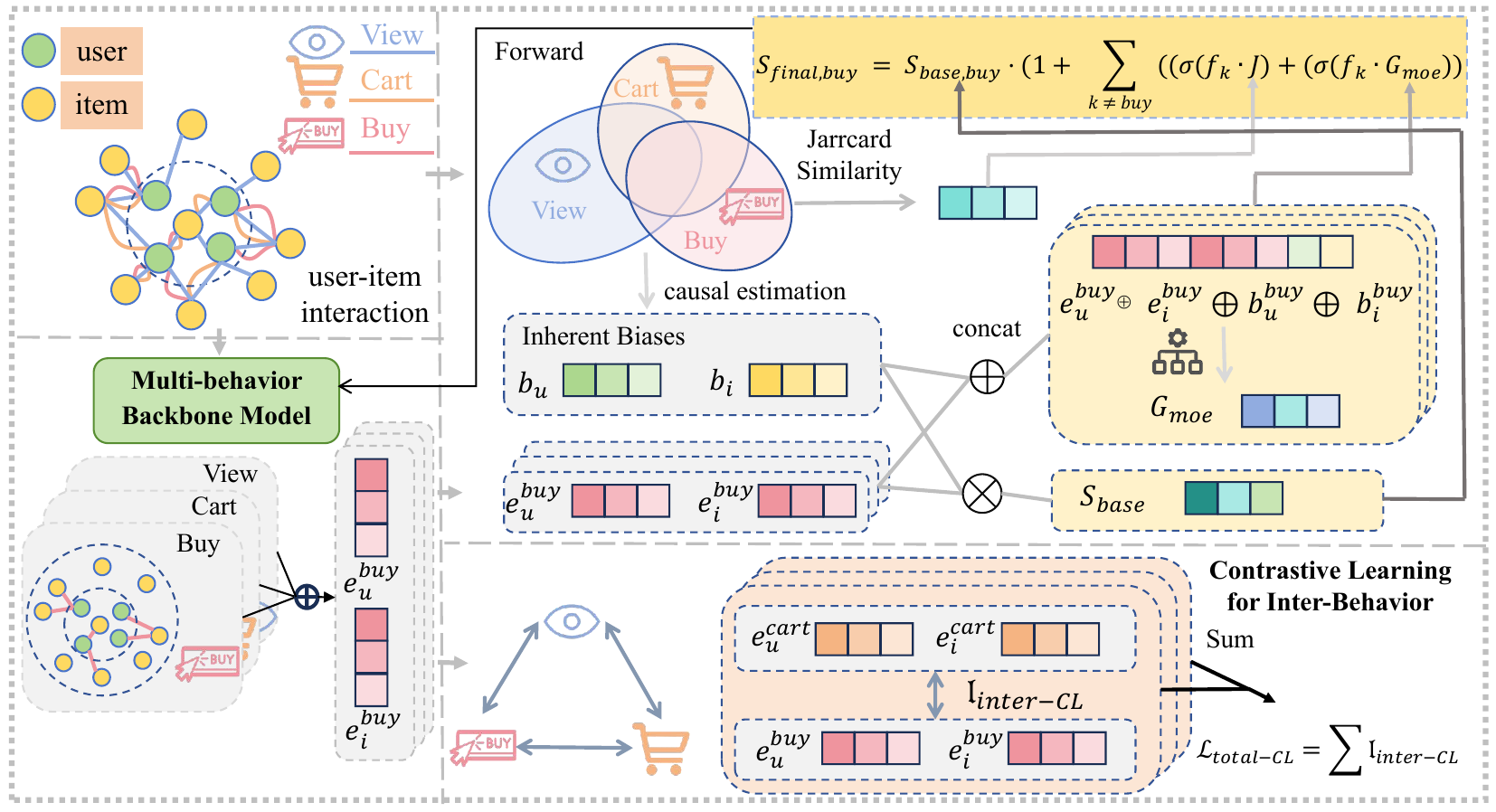}
    \caption{The overall framework of the MCLMR model}
    \label{fig:framework}
\end{figure*}

\paragraph{Problem Formulation}
The core challenge is multi-behavior bias, a confounder obscuring true preferences. This bias arises from two sources: (1) unique user interaction habits, where users exhibit distinct patterns, and (2) varying item behavior distributions, where items attract different engagement types. These biases create spurious correlations. Our goal is to disentangle the true preference signal from this confounding noise. Formally:

\begin{itemize}
    \item \textbf{Input:} The interaction graphs $\mathcal{G}$, the multi-behavior bias sets $\{\mathbf{b}_{u,k}, \mathbf{b}_{i,k}\}$, and the backbone embeddings $\{\mathbf{e}_{u,k}, \mathbf{e}_{i,k}\}$.
    \item \textbf{Objective:} Learn an enhancement function $f(\cdot)$ that performs causal debiasing and information fusion to accurately predict the score $\hat{y}_{ui}$ for the target behavior $K_t$:
    \begin{equation}
    \hat{y}_{ui} = f(u, i, \{\mathbf{e}_{u,k}, \mathbf{e}_{i,k}\}_{k=1}^K).
    \end{equation}
\end{itemize}

%% file: LaTeX/chapters/chapter_4_METHOD.tex
\section{Methodology}

In this paper, we propose \textbf{MCLMR}, a model-agnostic framework designed to enhance the performance of any backbone multi-behavior recommendation model. The architecture consists of three main modules: \textbf{Causal Preference Estimation} for debiased representations, \textbf{Adaptive Aggregation} for fusing cross-behavior signals, and \textbf{Bias-aware Contrastive Learning} for aligning representations. These modules constitute the training process, while the final prediction relies on a simple inner product in Section~\ref{subsec:44}.

\subsection{Causal Preference Estimation}
To elicit true, unbiased user preferences from data fraught with noise and inherent biases, we adopt a causal inference perspective to disentangle spurious correlations by explicitly modeling the data.

\begin{figure}[t!]
    \centering
    \includegraphics[width=0.9\columnwidth]{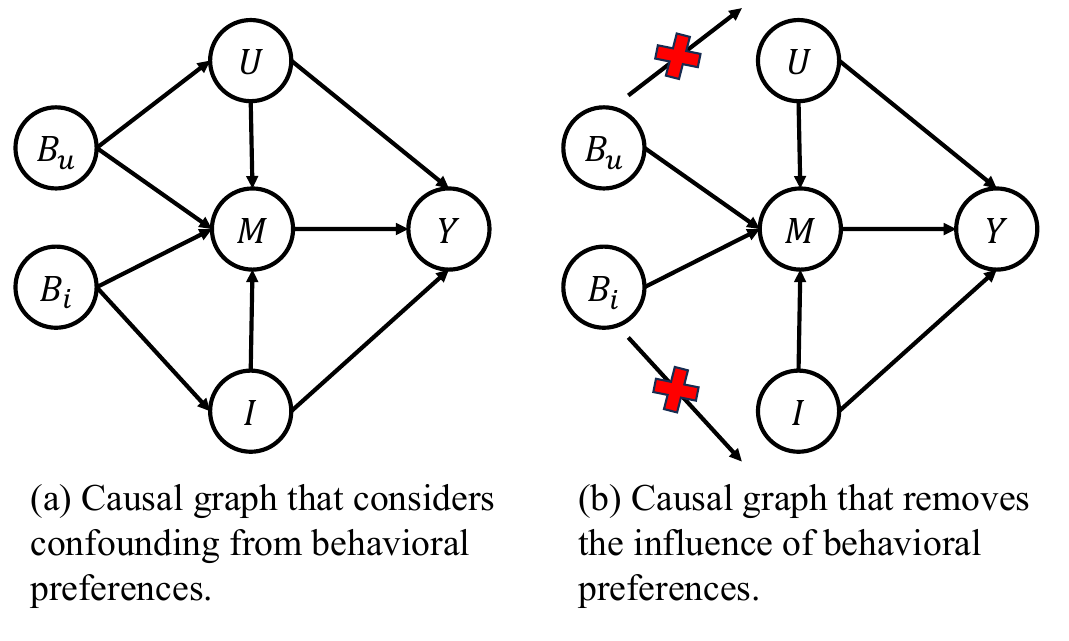}
    \caption{The proposed causal graph to illustrate the relationships among behavioral bias, true user preference, and observed interactions.}
    \label{fig:causal_graph}

\end{figure}

\subsubsection{Causal View of Multi-behavior Bias}
We construct a causal graph (Figure~\ref{fig:causal_graph}) to identify potential confounders. To facilitate this analysis, we formally define the variables and causal dependencies represented in the graph as follows:
\begin{itemize}
    \item $\mathbf{U}, \mathbf{I}$: Feature vectors for users and items, respectively.
    \item $\mathbf{M}$ (Mediating Behavior): The initial interaction, determined by feature matching ($\{U,I\}$) and inherent biases (user preference $B_u$, item tendency $B_i$).
    \item $\mathbf{Y}$ (Behavior Outcome): The final occurrence probability. Our goal is to predict the target outcome $Y_{K_t}$ of the behavior $K_t$.
    \item $\mathbf{B_u}, \mathbf{B_i}$: Represent the user's and item's  inherent multi-behavior biases, acting as hidden confounding variables.
\end{itemize}
The causal relationships identify $B_u$ and $B_i$ as confounders that create backdoor paths (e.g., $U \leftarrow B_u \rightarrow M \rightarrow Y$). Our objective is to block these paths via causal intervention.

\subsubsection*{Causal Effect Derivation}
To estimate the true causal effect, we apply the backdoor adjustment formula based on Pearl's do-calculus~\cite{10.5555/1642718}. The interventional probability $P(Y\,|\,\mathrm{do}(I), \mathrm{do}(U))$ is given by:
{\small
\begin{equation}\small
\label{equ:P_do_IU}
\begin{split}
    & P(Y|\operatorname{do}(I), \operatorname{do}(U)) = \sum_{B_{i}}P(B_{i})\sum_{B_{u}}P(B_{u}) \\
    &  \times \sum_{m}P(Y|M=m,I,U) 
     P(M=m|B_{i},B_{u},I,U).
\end{split}
\end{equation}}

Detailed derivation of Eq.~\ref{equ:P_do_IU} is provided in the appendix~\ref{sec:causal_effect}.

\subsubsection{Parameterizing Model Components and Biases}
\label{sec:parameterizing}

To make Eq.~\ref{equ:P_do_IU} tractable, we parameterize the key components, including the multi-behavior biases and the associated probability distributions, and map them to the actual computations within our neural network.

\paragraph{Bias Proxies via Generalized Behavioral Conversion Rate}


Using absolute counts can be misleading. To capture intrinsic propensities, we define bias proxies as relative frequencies. The User multi-behavior Bias ($b_{u,k}$) and Item multi-behavior Bias ($b_{i,k}$) are:
\begin{equation}
b_{u,k} = \frac{|\mathcal{I}_{u}^{k}|}{\sum_{j=1}^{K} |\mathcal{I}_{u}^{j}|},
\qquad
b_{i,k} = \frac{|\mathcal{U}_{i}^{k}|}{\sum_{j=1}^{K} |\mathcal{U}_{i}^{j}|}.
\end{equation}
While identifying an optimal proxy for latent confounders remains a widespread challenge in causal debiasing, our approach follows established conventions in the field (e.g., DecRS, iDCF, PDA). We utilize relative frequency as a practical and effective approximation to capture behavioral propensities, though we acknowledge that exploring more complex proxy formulations remains a valuable direction for future work.

Rather than treating these values as simple heuristics, we interpret them as computable proxies for the confounding variables ($B_u, B_i$) that govern interaction choices. Specifically, $b_{u,k}$ reflects a user's causal baseline propensity to `convert' a general state of engagement into behavior $k$. Unlike traditional conversion rates that model pairwise sequential transitions (e.g., view-to-click) with $O(K^2)$ complexity, our \textit{holistic conversion} approach efficiently captures global behavioral tendencies. This formulation provides a rigorous basis for causal adjustment while maintaining efficiency.

\paragraph{Parameterizing Probability Distributions}
We approximate the probability terms in the backdoor adjustment (Eq.~\ref{equ:P_do_IU}) using the defined components. The initial interaction propensity $P(M_k = 1\,|\,\cdot)$ is modeled as a debiased base score $S_{\text{base},k}$, combining the true matching degree $f_k(u,i)$ and the bias effect $g_k(b_{u,k}, b_{i,k})$:

\begin{equation}\footnotesize
P(M_k=1|B_{i,k},B_{u,k},U,I) \approx  S_{base,k}(u,i)  = f_{k}(u,i) \cdot g_k(b_{u,k},b_{i,k}).
\end{equation}
Subsequently, the outcome probability $P(Y_j = 1\,|\,M_k = 1, \dots)$ describes the likelihood of a final behavior $j$ occurring given the initial interaction. This is modulated by auxiliary behaviors via a cross-behavior contribution factor $S_{\text{contrib}}$:

\begin{equation}\footnotesize
\!P(Y_j\!=\! 1|M_k\!=\! 1, u, i) \!\approx\! f_{j}(u,i) \!\cdot\! S_{contrib}(M_k \!\!\to\!\! Y_j)(u,i).
\end{equation}
These parameterizations enable the model to approximate the interventional objective in a computationally tractable manner, laying the foundation for behavior-aware predictions.

\subsection{Adaptive Aggregation module}

The aggregation mechanism described below is a general framework applicable to any behavior $k$. For clarity and alignment with our main goal, we illustrate the aggregation mechanism for the target behavior $K_t$. The training prediction score, $S_{\text{final},K_t}(u,i)$, combines the base score with contributions from auxiliary behaviors:
\begin{equation}\small
S_{\text{final},K_t}(u,i) = S_{base,K_t}(u,i) \cdot \left(1 + C_{K_t}(u,i,b_{u,k},b_{i,k})\right).
\end{equation}
Here, the term $1$ represents the self-contribution from $K_t$, while the summation aggregates influence from auxiliary behaviors.
\begin{equation}\small
C_{K_t}(u,i,b_u,b_i) = \sum_{k \neq K_t} \operatorname{Con}(k \to K_t)(u,i,b_{u,k},b_{i,k}).
\end{equation}

\subsubsection*{Adaptive Aggregation for Training Objective}
The contribution $\operatorname{Con}(k \to K_t)$ is derived from the auxiliary matching degree $f_k(u,i)$, which is dynamically modulated by structural and semantic relevance via a dual-path gating mechanism:
\begin{equation}
\begin{split}\footnotesize
\nonumber \operatorname{Con}(k\! &\to\! K_t)(u,i,b_{u,k},b_{i,k}) 
 \!=\! \lambda_J \!\cdot\! \sigma(f_{k}(u,i) \!\cdot\! J(u,k,K_t)) \\
& + \lambda_M \cdot \sigma(f_{k}(u,i) \cdot g_{\text{moe}}(u,i,k,b_{u,k},b_{i,k})).
\end{split}
\end{equation}

where $\lambda_J$ and $\lambda_M$ are learnable parameters that balance the influence of the two paths. Next, we detail the estimation of the components in each path.

\paragraph{Structural Gating with Jaccard Similarity.}
This path's contribution relies on two elements: the true matching degree of the auxiliary behavior, $f_k(u,i)$, and its structural similarity to the target, $J(u,k,K_t)$. The Jaccard similarity quantifies the overlap of items the user $u$ interacted with under behavior $k$ and $K_t$.

\begin{equation}\small
J(u,k,K_t) = \frac{|\mathcal{I}_u^{k} \cap \mathcal{I}_u^{K_t}|}{|\mathcal{I}_u^{k} \cup \mathcal{I}_u^{K_t}| + \epsilon}.
\end{equation}
Here, $\mathcal{I}_u^{k}$ denotes the set of items with which user $u$ has interacted under behavior $k$. By multiplying $f_k$ with this structural score, we ensure that an auxiliary behavior contributes more significantly only if it is both intrinsically preferred and structurally relevant to the target.

\paragraph{Semantic Gating with Mixture-of-Experts (MoE).}
This path's contribution is also weighted by a semantic relevance score, $g_{\text{moe}}(u,i,k)$, from a MoE gating network. The gate dynamically assesses an auxiliary behavior's relevance by processing its corresponding user embeddings, item embeddings, and biases as input.

\begin{equation}
\begin{split}
&\text{input}_{\text{gate}} = [ \mathbf{e}_{u,k}, \mathbf{e}_{i,k}, \mathbf{b}_{u,k}, \mathbf{b}_{i,k}], \\
&g_{\text{moe}}(u,i,k,b_{u,k},b_{i,k}) = G_{\text{moe}}(\text{input}_{\text{gate}}).
\end{split}
\end{equation}
This design leverages an MoE gate to assess the semantic value of an auxiliary interaction; it uses the embeddings and biases of a specific behavior, k, to determine if the interaction is a strong signal or noise, thereby uncovering its deep, latent dependencies.

\subsection{Bias-aware Contrastive Learning module}
\label{sec:contrastive_learning}
To bridge the semantic gaps between behavior-specific representations, we introduce an adaptive contrastive learning framework. This module aligns the representation spaces across different actions, creating a more holistic and consistent user/item profile. Crucially, it mitigates data sparsity in key downstream behaviors (e.g., purchase) by transferring knowledge from denser, upstream interactions (e.g., view).

\subsubsection{Constructing User and Item-centric Contrastive Views}
We establish relational alignment from both user and item perspectives, treating each behavior type $k$ as an independent ``view''.

For any user $u$, their embeddings from two different behavioral views (e.g., $\mathbf{e}_u^{\text{view}}$, $\mathbf{e}_u^{\text{purchase}}$) are considered semantically linked and form a Positive Pair. Conversely, embeddings from different users within the same batch (e.g., $\mathbf{e}_u^{\text{view}}$, $\mathbf{e}_v^{\text{purchase}}$) are considered unrelated and form Negative Pairs.

Symmetrically, for any item $i$, its embeddings from two different behavioral views (e.g., $\mathbf{e}_i^{\text{view}}$, $\mathbf{e}_i^{\text{purchase}}$) constitute a Positive Pair. Embeddings from different items (e.g., $\mathbf{e}_i^{\text{view}}$, $\mathbf{e}_j^{\text{purchase}}$) form Negative Pairs. This encourages the model to learn behavior-invariant item characteristics.

\subsubsection{Adaptive Contrastive Objective with Personalized Bias-aware Temperature}


We adopt InfoNCE as the contrastive objective. Unlike fixed parameters that ignore heterogeneity, we introduce a personalized \textit{Bias-aware} adaptive temperature. The core intuition is that strong behavioral patterns (e.g., frequent purchases) yield reliable signals, warranting a lower $\tau$ to enforce stricter alignment. Conversely, sparse or noisy interactions require a higher $\tau$ to prevent overfitting.

Leveraging the cascading nature of multi-behavior data (e.g., view $\to$ purchase), we define reliability based on the \textit{downstream} behavior's bias. Consequently, the temperature $\tau$ for aligning views $k_i \to k_j$ is dynamically computed for each user $u$ as:
\begin{equation}
\label{tau}
\tau_{u, k_j} = \tau_0 \cdot (1 - \alpha \cdot b_{u, k_j}).
\end{equation}
where $\tau_0$ is a base temperature, $b_{u, k_j}$ is the user's bias for the downstream behavior $k_j$, and $\alpha \in [0, 1)$ is a scaling factor that controls the sensitivity of $\tau$ to the bias. An analogous formula is used for item-centric temperature $\tau_{i, k_j}$ using item bias $b_{i, k_j}$.

The final contrastive loss integrates both user and item-centric objectives. The loss for a user $u$ between views $k_i$ and $k_j$ is:
\begin{equation}\footnotesize
\mathcal{L}_{CL-U}(u_{k_i},u_{k_j}) = -\log \frac{\exp(\text{sim}(\mathbf{e}_u^{k_i}, \mathbf{e}_u^{k_j}) / \tau_{u, k_j})}{\sum_{v \in \mathcal{U}} \exp(\text{sim}(\mathbf{e}_u^{k_i}, \mathbf{e}_v^{k_j}) / \tau_{u, k_j})}.
\end{equation}
The total contrastive loss $\mathcal{L}_{CL}$ is the sum of all user-centric and item-centric losses over all pairs of distinct behaviors, balanced by a hyperparameter $\beta$:
\begin{equation}\small
\mathcal{L}_{CL} = \sum_{u, i \neq j} \mathcal{L}_{CL-U}(u_{k_i},u_{k_j}) + \beta \sum_{i, i' \neq j'} \mathcal{L}_{CL-I}(i_{k_{i'}},i_{k_{j'}}).
\end{equation}
By jointly optimizing this adaptive auxiliary loss, the model learns more robust, coherent, and representations that respect the varying reliability of interaction signals.
\subsection{Final Prediction for Recommendation}
\label{subsec:44}
At the inference stage, the final prediction score $\hat{y}_{ui}$ for the target behavior $K_t$ is determined by the simple inner product of the learned user and item embeddings:
\begin{equation}
\hat{y}_{ui} = \mathbf{e}_{u, K_t} \cdot \mathbf{e}_{i, K_t}^T
\end{equation}
This score serves as an unbiased estimate of the true user preference, as confounding biases were addressed during the model's causal training phase. It is then used to rank items for recommendation.

%% file: LaTeX/chapters/chapter_5_EXPERIMENTS.tex
\section{Experiments}
In this section, we conduct comprehensive experiments to evaluate MCLMR's effectiveness, examining its performance improvements across different models and datasets (RQ1), the impact of key components (RQ2), hyperparameter sensitivity (RQ3), the effectiveness of causal deconfounding (RQ4), and computational efficiency (RQ5).



\subsection{Experimental Settings}


\subsubsection{Datasets.}
We evaluate our proposed method on three real-world public benchmark datasets. The pre-processing of these datasets is consistent with the previous methods~\cite{cai2025neighborhood,yin2024hecgcnhypergraphenhancedcascading}. The statistics of these datasets are summarized in Table~\ref{tab:dataset_stats}.

\begin{table}[t!]
\centering
\caption{Statistics of the experimental datasets.}
\label{tab:dataset_stats}

\resizebox{0.9\columnwidth}{!}{%
\begin{tabular}{lrrrrrr}
\toprule
\textbf{Dataset} & \textbf{Users} & \textbf{Items} & \textbf{view} & \textbf{collect} & \textbf{cart} & \textbf{buy} \\
\midrule
Tmall            & 41,738           & 11,953           & 1,813,498 & 221,514 & 1,996 & 255,586 \\
Jdata            & 93,334           & 24,624           & 1,681,430 & 45,613 & 49,891 & 333,383 \\
Taobao         & 15,449           & 11,953           & 873,954 &\multicolumn{1}{c}{-}& 195,476 & 92,180  \\

\bottomrule
\end{tabular}%
}

\end{table}

\subsubsection{Baselines.}
\begin{table*}[t!]
\centering
\caption{Performance Comparison with Various Methods (The best performance is highlighted in \textbf{bold}). The statistical significance of improvements over the best baseline is determined by a paired t-test. Symbols $^{*}$, $^{**}$, and $^{***}$ denote statistical significance at the level of $p < 0.05$, $p < 0.01$, and $p < 0.001$, respectively.}
\label{tab:main_results}
{
\renewcommand{\arraystretch}{1.0}
\resizebox{\textwidth}{!}{%
\begin{tabular}{c|l|cccc|cccc|cccc}
\toprule
\multirow{2}{*}{\textbf{Type}} & \multirow{2}{*}{\textbf{Model}} & \multicolumn{4}{c|}{\textbf{Tmall}} & \multicolumn{4}{c|}{\textbf{Jdata}} & \multicolumn{4}{c}{\textbf{Taobao}} \\
\cmidrule{3-14}
& & H@10 & N@10 & H@50 & N@50 & H@10 & N@10 & H@50 & N@50 & H@10 & N@10 & H@50 & N@50 \\
\midrule
\multirow{4}{*}{\parbox{3cm}{\centering\textbf{Single-behavior Baselines}}} & MF-BPR & 0.0230 & 0.0124 & 0.0434 & 0.0166 & 0.1850 & 0.1238 & 0.2652 & 0.1417 & 0.0076 & 0.0036 & 0.0151 & 0.0052 \\
& NCF & 0.0301 & 0.0153 & 0.0678 & 0.0231 & 0.2090 & 0.1410 & 0.2934 & 0.1599 & 0.0236 & 0.0128 & 0.0483 & 0.0175 \\
\cmidrule{2-14}
& LightGCN & 0.0463 & 0.0226 & 0.1369 & 0.0414 & 0.2932 & 0.1697 & 0.5374 & 0.2246 & 0.0156 & 0.0077 & 0.0483 & 0.0147 \\
& \textbf{LightGCN+MCLMR} & 0.1142 & 0.0583 & 0.2729 & 0.0920 & 0.5026 & 0.2914 & 0.7537 & 0.3503 & 0.1142 & 0.0598 & 0.2815 & 0.0962 \\
\midrule
\multirow{7}{*}{\parbox{3cm}{\centering\textbf{Multi-behavior Baselines}}} & RGCN & 0.0316 & 0.0157 & 0.0826 & 0.0262 & 0.2406 & 0.1444 & 0.4873 & 0.1891 & 0.0215 & 0.0104 & 0.0439 & 0.0141 \\
& NMTR & 0.0517 & 0.0250 & 0.1498 & 0.0456 & 0.3142 & 0.1717 & 0.5227 & 0.2198 & 0.0282 & 0.0137 & 0.0578 & 0.0193 \\
& MBGCN & 0.0549 & 0.0285 & 0.1285 & 0.0438 & 0.2803 & 0.1572 & 0.5045 & 0.1984 & 0.0509 & 0.0294 & 0.1009 & 0.0415\\
& GNMR & 0.0393 & 0.0193 & 0.1071 & 0.0332 & 0.3068 & 0.1581 & 0.4607 & 0.2029 & 0.0368 & 0.0216 & 0.0735 & 0.0306 \\
& S-MBRec & 0.0694 & 0.0362 & 0.1553 & 0.0544 & 0.4125 & 0.2779 & 0.6036 & 0.3203 & 0.1185 & 0.0659 & 0.2405 & 0.0901 \\
& MB-HGCN & 0.1461 & 0.0770 & 0.3149 & 0.1130 & 0.5338 & 0.3238 & 0.7749 & 0.3804 & 0.1380 & 0.0728 & 0.2751 & 0.1025 \\
& NSED & 0.1480 & 0.0817 & 0.2814 &  0.1102 & 0.5519 & 0.3378 & 0.7392 & 0.3820 & 0.1652 & 0.0912 & 0.2957 & 0.1198 \\
\midrule
\multirow{2}{*}{\parbox{3cm}{\centering\textbf{Integrated Causal Multi-behavior}}}
& CVID & 0.1542 & 0.0829 & 0.3236 & 0.1215 & 0.4258 & 0.2424 & 0.6685 & 0.2979 & 0.1424 & 0.0768 & 0.2862 & 0.1087 \\
& CMSR & 0.1585 & 0.0877 & 0.3089 & 0.1206 & 0.5117 & 0.3095 & 0.7276 & 0.3713 & 0.1295 & 0.0713 & 0.2634 & 0.0989 \\
\midrule
\multirow{6}{*}{\parbox{3cm}{\centering\textbf{Multi-behavior + MCLMR}}}& CRGCN & 0.0840 & 0.0442 & 0.1994 & 0.0685 & 0.4234 & 0.2479 & 0.6749 & 0.3064 & 0.0882 & 0.0496 & 0.1962 & 0.0730 \\
& \textbf{CRGCN+MCLMR} & $0.1449^{***}$ & $0.0803^{***}$ & $0.2725^{**}$ & $0.1075^{***}$ & $0.5003^{***}$ & $0.3054^{**}$ & $0.7007^{***}$ & $0.3521^{***}$ & $0.1464^{**}$ & $0.0839^{***}$ & $0.2753^{***}$ & $0.1122^{***}$ \\
\cmidrule{2-14}
& BCIPM & 0.1411 & 0.0741 & 0.3057 & 0.1093 & 0.3851 & 0.2158 & 0.6264 & 0.2711 & 0.1093 & 0.0596 & 0.2462 & 0.0894 \\
& \textbf{BCIPM+MCLMR} & 0.1755 & 0.0970 & 0.3320 & 0.1306 & 0.5575 & 0.3381 & 0.7655 & 0.3875 & 0.1635 & 0.0926 & 0.3023 & 0.1232 \\
\cmidrule{2-14}
& HEC-GCN & 0.1670 & 0.0912 & 0.3131 & 0.1226 & 0.5463 & 0.3360 & 0.7426 & 0.3826 & 0.1807 & 0.0980 & 0.3439 & 0.1341 \\
& \parbox{3cm}{\textbf{HEC-GCN+MCLMR}} & $\textbf{0.1896}^{**}$ & $\textbf{0.1041}^{**}$ & $\textbf{0.3400}^{**}$ & $\textbf{0.1367}^{**}$ & $\textbf{0.5744}^{***}$ & $\textbf{0.3548}^{**}$ & $\textbf{0.7622}^{***}$ & $\textbf{0.3996}^{***}$ & $\textbf{0.1912}^{**}$ & $\textbf{0.1055}^{***}$ & $\textbf{0.3575}^{*}$ & $\textbf{0.1379}^{*}$ \\
\bottomrule
\end{tabular}%
}
}

\end{table*}

\begin{table*}[t!]
\centering
\caption{Performance Comparison with Pluggable Causal Models (The best results for each backbone are in \textbf{bold}).}
\label{tab:causal_results_full}
\resizebox{\textwidth}{!}{%
\begin{tabular}{l|l|cccc|cccc|cccc}
\toprule
\multirow{2}{*}{\textbf{Base Model}} & \multirow{2}{*}{\textbf{Method}} & \multicolumn{4}{c|}{\textbf{Tmall}} & \multicolumn{4}{c|}{\textbf{Jdata}} & \multicolumn{4}{c}{\textbf{Taobao}} \\
\cmidrule(l){3-14}
& & H@10 & N@10 & H@50 & N@50 & H@10 & N@10 & H@50 & N@50 & H@10 & N@10 & H@50 & N@50 \\
\midrule
\multirow{6}{*}{\parbox{3cm}{\centering{\textbf{LightGCN}}}} 
& \parbox{3cm}{Original} & 0.0463 & 0.0226 & 0.1369 & 0.0414 & 0.2932 & 0.1697 & 0.5374 & 0.2246 & 0.0156 & 0.0077 & 0.0483 & 0.0147 \\
& + SNIPS & 0.0478 & 0.0284 & 0.1533 & 0.0556 & 0.4149 & 0.2438 & 0.6948 & 0.3067 & 0.0141 & 0.0073 & 0.0427 & 0.0134 \\
& + CausE & 0.0390 & 0.0188 & 0.1184 & 0.0351 & 0.3005 & 0.1766 & 0.5654 & 0.2350 & 0.0185 & 0.0089 & 0.0528 & 0.0161 \\
& + Multi-IPW & 0.0573 & 0.0322 & 0.1738 & 0.0657 & 0.4150 & 0.2520 & 0.6661 & 0.3084 & 0.0241 & 0.0136 & 0.0745 & 0.0241 \\
& + DCCL & 0.0952 & 0.0497 & 0.2386 & 0.0799 & 0.4822 & 0.2886 & 0.7347 & 0.3472 & 0.1067 & 0.0553 & 0.2629 & 0.0890 \\
& + MCLMR & \textbf{0.1142} & \textbf{0.0583} & \textbf{0.2729} & \textbf{0.0920} & \textbf{0.5026} & \textbf{0.2914} & \textbf{0.7537} & \textbf{0.3503} & \textbf{0.1142} & \textbf{0.0598} & \textbf{0.2815} & \textbf{0.0962} \\
\midrule
\multirow{6}{*}{\parbox{3cm}{\centering{\textbf{CRGCN}}}} 
& Original & 0.0840 & 0.0442 & 0.1994 & 0.0685 & 0.4234 & 0.2479 & 0.6749 & 0.3064 & 0.0882 & 0.0496 & 0.1962 & 0.0730 \\
& + SNIPS & 0.1197 & 0.0654 & 0.2615 & 0.0955 & 0.4283 & 0.2456 & 0.6788 & 0.3041 & 0.1082 & 0.0635 & 0.2205 & 0.0879 \\
& + CausE & 0.1185 & 0.0632 & 0.2608 & 0.0933 & 0.3828 & 0.2161 & 0.6299 & 0.2727 & 0.0917 & 0.0518 & 0.1998 & 0.0752 \\
& + Multi-IPW & 0.1201 & 0.0654 & 0.2590 & 0.0950 & 0.4189 & 0.2432 & 0.6722 & 0.3024 & 0.1080 & 0.0633 & 0.2207 & 0.0878 \\
& + DCCL & 0.1144 & 0.0624 & 0.2381 & 0.0886 & 0.4699 & 0.2860 & 0.6833 & 0.3353 & 0.1174 & 0.0667 & 0.2432 & 0.0942 \\
& + MCLMR & \textbf{0.1449} & \textbf{0.0803} & \textbf{0.2725} & \textbf{0.1075} & \textbf{0.5003} & \textbf{0.3054} & \textbf{0.7007} & \textbf{0.3521} & \textbf{0.1464} & \textbf{0.0839} & \textbf{0.2753} & \textbf{0.1122} \\
\midrule
\multirow{6}{*}{\parbox{3cm}{\centering{\textbf{HEC-GCN}}}}
& Original & 0.1670 & 0.0912 & 0.3131 & 0.1226 & 0.5463 & 0.3360 & 0.7426 & 0.3826 & 0.1807 & 0.0980 & 0.3439 & 0.1341 \\
& + SNIPS & 0.1744 & 0.0950 & 0.3278 & 0.1279 & 0.5156 & 0.3148 & 0.7080 & 0.3597 & 0.1851 & 0.1027 & 0.3345 & 0.1358 \\
& + CausE & 0.1768 & 0.0961 & 0.3365 & 0.1308 & 0.5568 & 0.3396 & 0.7459 & 0.3842 & 0.1847 & 0.1005 & 0.3340 & 0.1334 \\
& + Multi-IPW & 0.1523 & 0.0824 & 0.3047 & 0.115 & 0.5275 & 0.3167 & 0.7548 & 0.3701 & 0.1584 & 0.0862 & 0.3312 & 0.1241 \\
& + DCCL & 0.1759 & 0.0967 & 0.3237 & 0.1286 & 0.5346 & 0.3250 & 0.7306 & 0.3711 & 0.1863 & 0.1034 & 0.3259 & 0.1342 \\
& + MCLMR & \textbf{0.1896} & \textbf{0.1041} & \textbf{0.3400} & \textbf{0.1367} & \textbf{0.5744} & \textbf{0.3548} & \textbf{0.7622} & \textbf{0.3996} & \textbf{0.1912} & \textbf{0.1055} & \textbf{0.3575} & \textbf{0.1379} \\
\bottomrule
\end{tabular}%
}
\end{table*}
To validate the effectiveness of MCLMR, we compared it with numerous baseline models in recent years, which can be divided into three categories: (1) Single-behavior methods: MF-BPR~\cite{bpr}, NCF~\cite{ncf} and LightGCN~\cite{lightgcn}, (2) Multi-behavior methods: RGCN~\cite{rgcn}, NMTR~\cite{nmtr}, MBGCN~\cite{mbgcn}, GNMR~\cite{gnmr}, S-MBRec~\cite{smbrec}, MB-HGCN~\cite{mbhgcn}, CRGCN~\cite{crgcn}, BCIPM~\cite{bipn}, HEC-GCN~\cite{yin2024hecgcnhypergraphenhancedcascading}.(3) Integrated Causal Multi-behavior methods: CVID~\cite{10.1145/3745023}, CMSR~\cite{cmsr}.(4) Pluggable Causal methods: SNIPS~\cite{NIPS2015_39027dfa}, CausE~\cite{10.1145/3240323.3240360}, Multi-IPW~\cite{DBLP:journals/corr/abs-1910-09337}, DCCL~\cite{10.1145/3543873.3584637}.
We denote the models enhanced by our framework as \textbf{LightGCN+MCLMR}, \textbf{HEC-GCN+MCLMR}, \textbf{CRGCN+MCLMR}, and \textbf{BCIPM+MCLMR}.

\subsubsection{Evaluation Metrics.}
To evaluate the performance of top-N recommendation, we adopt two widely used metrics: Hit Ratio (HR@K) and Normalized Discounted Cumulative Gain (NDCG@K). We set K to 10 and 50. The final prediction is made on the target behavior ('buy').

\subsubsection{Implementation Details.}

We implement all models in PyTorch on an NVIDIA RTX 4090 GPU using the Adam~\cite{adam} optimizer. For fair comparison, we adopt optimal backbone settings and fix the embedding size at 64. All MCLMR-specific hyperparameters were carefully tuned; detailed search spaces and final configurations are provided in Appendix~\ref{sec:hyperparameter}.


\subsection{Overall Performance Comparison (RQ1)}

To answer RQ1, tables~\ref{tab:main_results} and~\ref{tab:causal_results_full} summarize the comparison results. MCLMR consistently enhances a wide range of backbone models across all datasets, from single-behavior methods to advanced architectures. Notably, the enhancement enables simpler models to surpass complex baselines; for example, LightGCN+MCLMR outperforms CRGCN on Jdata in terms of NDCG@10 (0.2914 vs. 0.2479). Substantial improvements are also observed in strong, specialized models (e.g., HEC-GCN on Tmall). Furthermore, compared with other pluggable causal methods like SNIPS and DCCL, our framework achieves superior performance on identical backbones, validating its effectiveness and model-agnostic nature.

\subsection{Ablation Study (RQ2)}

To answer RQ2, We conduct a comprehensive ablation study using CRGCN and BCIPM on Tmall and Jdata to validate MCLMR's components. We compare the full model against six variants: (1) \textbf{w/o U-Bias}: Removes the debiasing module for user behavior habits. (2) \textbf{w/o I-Bias}: Removes the debiasing module for item multi-behavior bias. (3) \textbf{w/o Agg.}: Removes the entire adaptive information aggregation module. (4) \textbf{w/o Jaccard}: Removes the Jaccard gate within the aggregation module. (5) \textbf{w/o MoE}: Removes the Mixture-of-Experts (MoE) gate within the aggregation module. (6) \textbf{w/o CL}: Removes the contrastive learning module.

As detailed in Tables~\ref{tab:ablation_tmall} and~\ref{tab:ablation_jdata}, the full model consistently yields the best performance. The degradation across all variants confirms that every component is indispensable. Specifically, the significant drop in \textbf{w/o CL} identifies contrastive learning as a critical regularizer. Similarly, declines in \textbf{w/o Agg.} and \textbf{w/o U/I-Bias} validate the vital roles of adaptive aggregation in preventing negative transfer and causal debiasing in disentangling confounders.

\begin{table}[t!]
\centering
\caption{Ablation study results on the Tmall dataset. The best performance for each backbone is in \textbf{bold}.}
\label{tab:ablation_tmall}

\resizebox{0.95\columnwidth}{!}{%
\begin{tabular}{l|cc|cc}
\toprule
\multirow{2}{*}{\textbf{Model Variant}} & \multicolumn{2}{c|}{\textbf{BCIPM+MCLMR}} & \multicolumn{2}{c}{\textbf{CRGCN+MCLMR}} \\
\cmidrule{2-5}
& HR@10 & NDCG@10 & HR@10 & NDCG@10 \\
\midrule
w/o U-Bias & 0.1724 & 0.0949 & 0.1385 & 0.0763 \\
w/o I-Bias & 0.1716 & 0.0932 & 0.1380 & 0.0756 \\
w/o Jaccard & 0.1737 & 0.0946 & 0.1323 & 0.0718 \\
w/o MoE & 0.1719 & 0.0945 & 0.1400 & 0.0771 \\
w/o Agg. & 0.1675 & 0.0925 & 0.1309 & 0.0705 \\
w/o CL & 0.1476 & 0.0805 & 0.1213 & 0.0557 \\
\midrule
\textbf{Full Model} & \textbf{0.1755} & \textbf{0.0970} & \textbf{0.1449} & \textbf{0.0803} \\
\bottomrule
\end{tabular}
}

\end{table}

\begin{table}[t!]
\caption{Ablation study results on the Jdata dataset. The best performance for each backbone is in \textbf{bold}.}
\label{tab:ablation_jdata}
\centering
\resizebox{0.95\columnwidth}{!}{%
\begin{tabular}{l|cc|cc}
\toprule
\multirow{2}{*}{\textbf{Model Variant}} & \multicolumn{2}{c|}{\textbf{BCIPM+MCLMR}} & \multicolumn{2}{c}{\textbf{CRGCN+MCLMR}} \\
\cmidrule{2-5}
& HR@10 & NDCG@10 & HR@10 & NDCG@10 \\
\midrule
w/o U-Bias & 0.5431 & 0.3301 & 0.4518 & 0.2741 \\
w/o I-Bias & 0.5454 & 0.3320 & 0.4623 & 0.2797 \\
w/o Jaccard & 0.5539 & 0.3359 & 0.4694 & 0.2843 \\
w/o MoE & 0.5542 & 0.3361 & 0.4709 & 0.2865 \\
w/o Agg. & 0.5461 & 0.3328 & 0.4597 & 0.2770 \\
w/o CL & 0.4521 & 0.2660 & 0.4567 & 0.2522 \\
\midrule
\textbf{Full Model} & \textbf{0.5575} & \textbf{0.3381} & \textbf{0.5003} & \textbf{0.3054} \\
\bottomrule
\end{tabular}
}

\end{table}

\subsection{Parameter Analysis (RQ3)}
\begin{figure}[t!]
\centering
\includegraphics[width=0.50\textwidth]{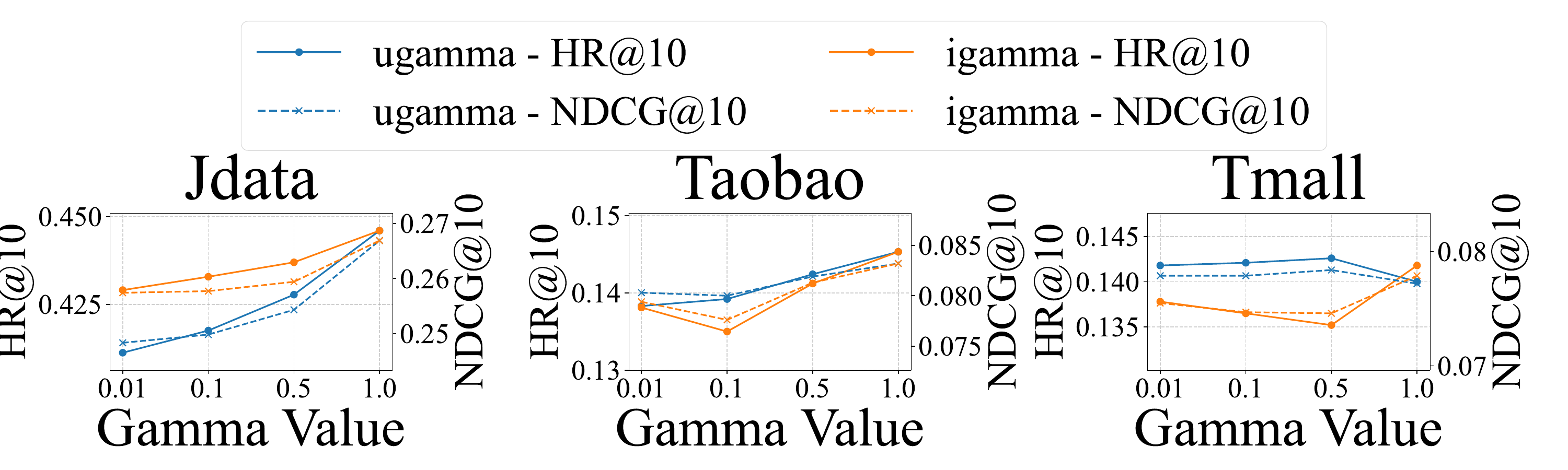} 
\caption{Impact of the debiasing coefficient \(\gamma\). Performance is shown for HR@10 (left) and NDCG@10 (right).}
\label{fig:gamma}
\end{figure}

\begin{figure}[t!]
\centering
\includegraphics[width=0.5\textwidth]{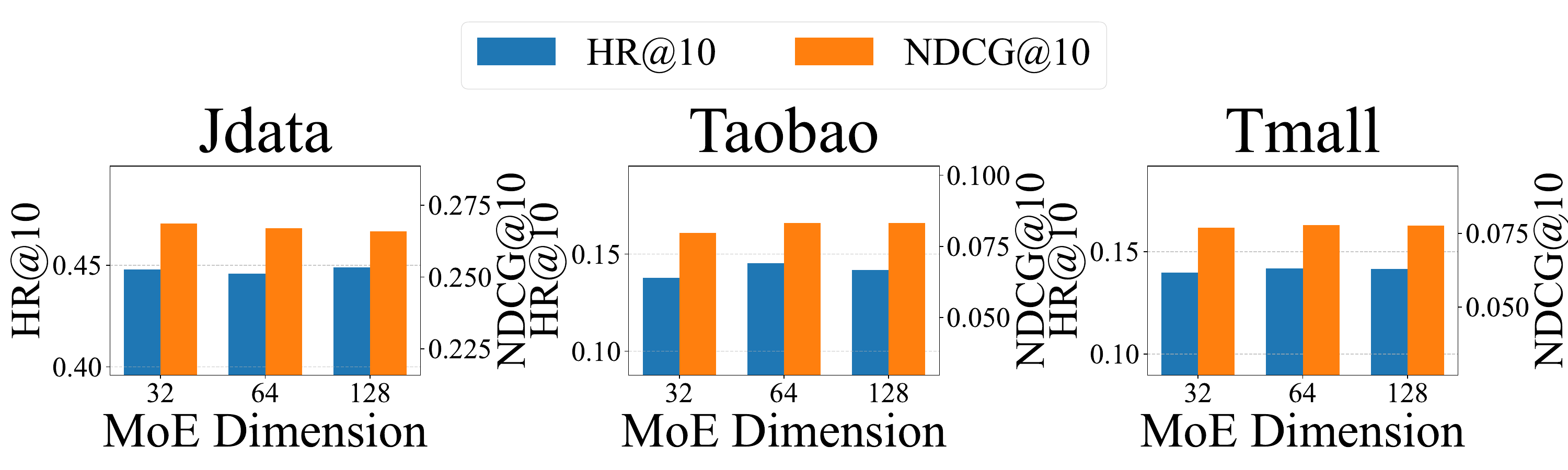} 
\caption{Impact of the MoE expert dimension. Performance is shown for HR@10 (left) and NDCG@10 (right). }
\label{fig:moe_dim}
\end{figure}

This subsection examines MCLMR's sensitivity to hyperparameters using the CRGCN backbone. Regarding the debiasing coefficient $\gamma$ (Fig.~\ref{fig:gamma}), a small value (0.01) is universally optimal for items, whereas the optimal user $\gamma$ is dataset-dependent (larger for Jdata, smaller for Taobao). For the MoE expert dimension (Fig.~\ref{fig:moe_dim}), performance peaks at 64, providing the optimal trade-off between expressiveness and generalization; larger capacities lead to overfitting.

\paragraph{Effectiveness of Bias-aware Temperature}
We investigate the temperature parameter $\tau$ in our contrastive loss. Unlike standard fixed or learnable approaches, we dynamically adapt $\tau$ to behavioral biases to balance the primary BPR and auxiliary contrastive tasks. Specifically, high-bias behaviors (dominating BPR) require distinct penalties compared to sparse ones. We validate this design against four variants (Fixed, Learnable, Random, Alternative Mappings), with detailed settings and comparisons provided in Appendix~\ref{app:temperature}.

\subsection{In-depth Analyses (RQ4)}

In this subsection, we conduct a detailed analysis using the CRGCN backbone to investigate three key questions (analyses on other backbones like HECGCN and LightGCN are deferred to Appendix~\ref{sec:appendix_analysis}). Specifically, we aim to determine: 1) the primary sources of performance improvements; 2) consistency across different user groups; and 3) effectiveness in recommending high-quality items.

\subsubsection{Improvements in Active and Less-active User Groups}
\begin{figure}[t!]
\centering
\includegraphics[width=0.5\textwidth]{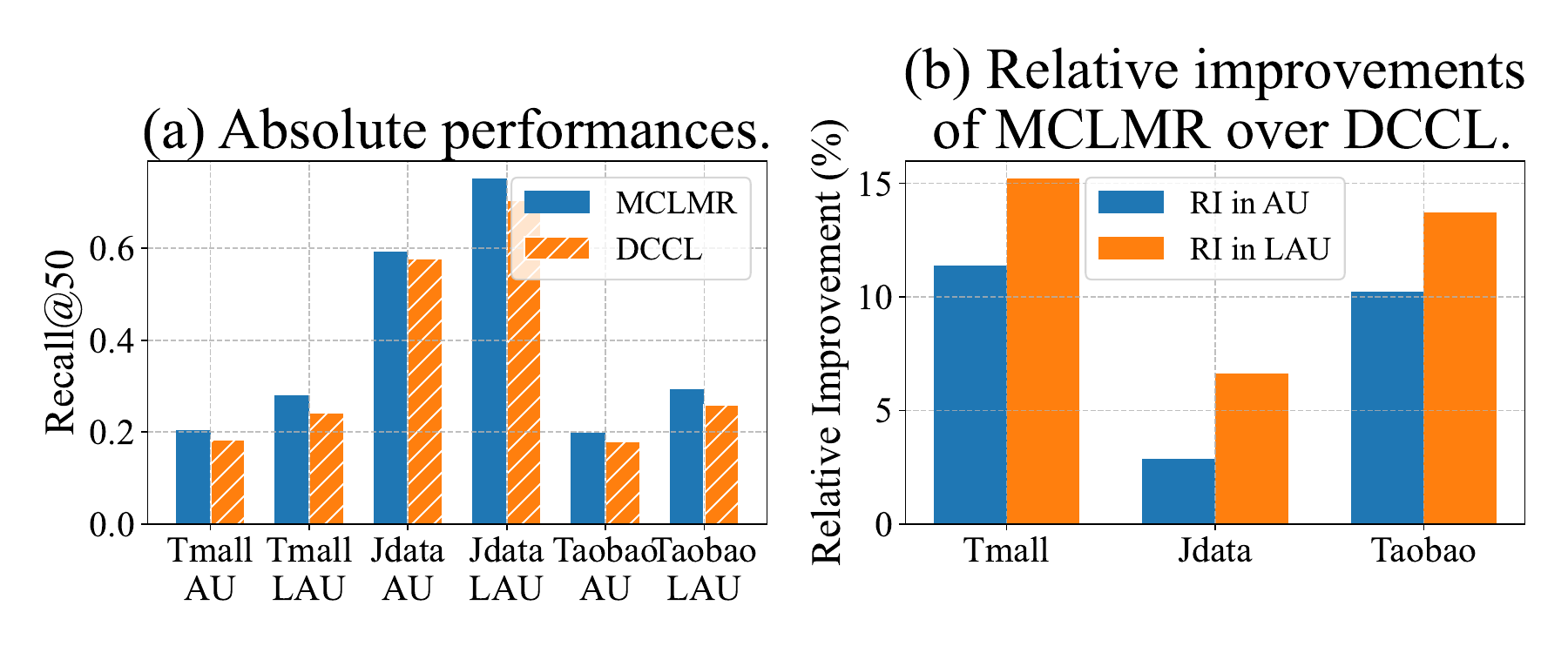} 
\caption{Performance of CRGCN+MCLMR and CRGCN+DCCL in active and less-active user groups. (a) the absolute performance; and (b) the relative improvements of MCLMR over DCCL. ”AU” and ”LAU” are short for the active user group and the less-active user group, respectively. In (a), bars with slash and without slash corresponds to DCCL and MCLMR, respectively. Better viewed in color.}
\label{fig:user}
\end{figure}
\begin{figure}[t!]
\centering
\includegraphics[width=0.5\textwidth]{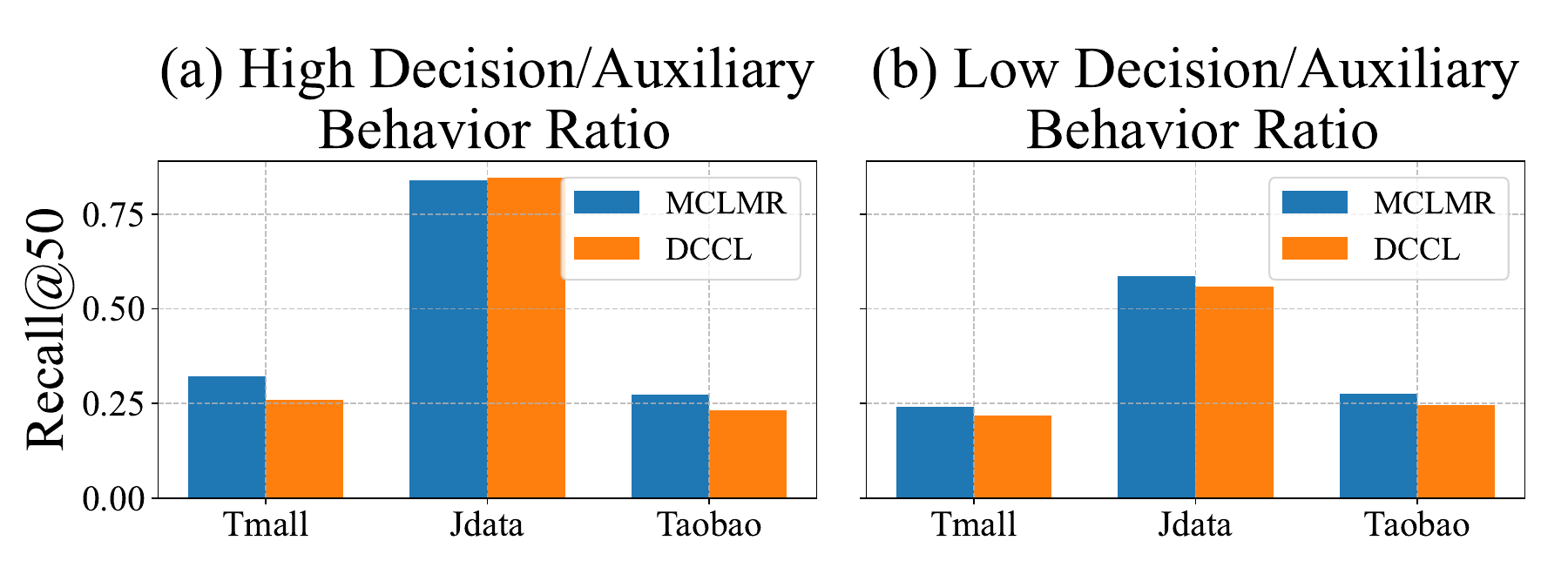} 
\caption{Recall@50 of CRGCN+MCLMR and CRGCN+DCCL in item groups with high and low Decision/Auxiliary Behavior ratio.}
\label{fig:item}
\end{figure}



We examine whether MCLMR benefits active or less-active user groups differently. For Tmall, Jdata, and Taobao, the top 20\% of users by interaction count are classified as \textit{active} (AU), and the rest as \textit{less-active} (LAU). We compare MCLMR and DCCL on: 1) absolute Recall@50 performance; and 2) relative improvements.

Figure~\ref{fig:user} summarizes the results. We observe that: 
1) Better Performance in LAU: As shown in Fig.~\ref{fig:user}(a), both models perform better in the LAU group across all datasets. 2) Overall Superiority: MCLMR consistently outperforms DCCL in both active and less-active groups. 3) Greater Relative Improvement in LAU: As shown in Fig.~\ref{fig:user}(b), the relative improvement (RI) of MCLMR over DCCL is notably higher in the LAU group. These results indicate MCLMR is particularly effective for long-tail users, helping alleviate the ``Matthew effect'' by ensuring less-active users receive high-quality recommendations despite sparse histories.
\subsubsection{Recommendation Quality w.r.t. Decision/Auxiliary Behavior Ratio}



Whether the proposed MCLMR can achieve consistent and unbiased user preference estimations is our concern. Thus, in this part, we sort items according to the \textit{decision/auxiliary behavior ratio} and divide them into two groups: the top 20\% of items with a high ratio, and the remaining 80\% with a low ratio. High-ratio items are considered higher quality as they align with target actions, whereas over-recommending low-ratio items may hurt user experience. We use recall for evaluation.

The performance on these groups is shown in Figure~\ref{fig:item}. In the high-ratio group (Figure~\ref{fig:item}(a)), MCLMR significantly outperforms DCCL on Tmall and Taobao, with comparable results on Jdata, demonstrating its capability to recommend high-quality items. Furthermore, in the low-ratio group (Figure~\ref{fig:item}(b)), MCLMR consistently leads across all datasets. These results suggest that MCLMR is more precise in recommending high-quality items and robust in handling low-quality ones. We attribute this to the removal of hidden confounders that might otherwise promote items with high auxiliary engagement but low decision value.

\subsection{Complexity Analysis (RQ5)}
To evaluate the computational overhead of our proposed MCLMR framework, we analyze its time complexity in comparison with three state-of-the-art multi-behavior recommendation models: CRGCN, HEC-GCN, and BCIPM.
\subsubsection{Notations}
We define the following symbols: $U$ and $I$ are the total number of users and items; $d$ is the embedding dimension; $B$ is the number of behaviors; $K$ is the number of GCN layers; $E_{\text{total}}$ is the total number of edges across all behaviors; $H$ is the number of hyperedges in the hypergraph; and $u_s$ is the number of unique users within a batch.
\subsubsection{Complexity of the MCLMR Framework}
Our MCLMR framework introduces an additional computational overhead, $\mathcal{O}_{\text{MCLMR}}$, which is composed of three main modules: (1) Jaccard Aggregation Module: $\mathcal{O}(N \cdot B^2 \cdot r)$. (2) MoE Gating Aggregation Module: $\mathcal{O}(N \cdot B^2 \cdot k \cdot d)$. (3) Multi-behavior Contrastive Learning Module: $\mathcal{O}((u_s + i_s) \cdot B^2 \cdot d^2)$.
Thus, the total incremental complexity is $\mathcal{O}_{\text{MCLMR}} = \mathcal{O}(N \cdot B^2 \cdot (r + k \cdot d) + (u_s + i_s) \cdot B^2 \cdot d^2)$.
\subsubsection{Complexity Comparison and Analysis}
Table~\ref{tab:complexity} summarizes the original dominant complexities of the backbone models and the total complexity after integrating MCLMR.

\begin{table}[t!]
\caption{Time complexity of baseline models before and after integration with the MCLMR framework.}
\label{tab:complexity}
\centering
\resizebox{\columnwidth}{!}{%
\begin{tabular}{l l l}
\toprule
\textbf{Model} & \textbf{Core Computational Modules} & \textbf{Dominant Time Complexity} \\
\midrule
HEC-GCN & \makecell{GCNs \& Hypergraph GCN \\ \& Attention \& Contrastive Learning}& $\mathcal{O}(L \cdot E_{\text{total}} \cdot d + K \cdot (M+N) \cdot S \cdot d + K \cdot (M^2 + N^2) \cdot d)$ \\
CRGCN & \makecell{Cascading GCN Propagation} & $\mathcal{O}(K \cdot E_{\text{total}} \cdot d)$ \\
BCIPM & \makecell{GCNs \& Neighborhood Aggregation} & $\mathcal{O}(K \cdot E_{\text{total}} \cdot d + \mathbf{b_s} \cdot \mathbf{L_{max}} \cdot \mathbf{d^2})$ \\
\hline
\textbf{+MCLMR} & \textit{+ Adaptive Aggregation + Bias-aware CL} & $ + \mathcal{O}(N \cdot B^2 \cdot (r + k \cdot d) + (u_s + i_s) \cdot B^2 \cdot d^2)$ \\
\bottomrule
\end{tabular}%
}
\end{table}

The analysis confirms that MCLMR introduces a fixed computational overhead, demonstrating its model-agnostic nature. The relative impact of this overhead is more significant on simpler models like CRGCN but less pronounced on inherently complex ones such as HEC-GCN. Notably, for BCIPM, our framework adds a complexity term that scales with the square of the batch size ($u_s^2$), increasing the model's sensitivity to this hyperparameter.
\paragraph{Empirical Runtime and Latency}
In addition to the theoretical complexity analysis, we evaluated the actual training overhead and online inference latency on real-world GPUs to demonstrate its practical efficiency. The empirical results confirm that MCLMR introduces only a minor increase in training time (e.g., approx. 4\% increase for HEC-GCN) while maintaining the same inference speed as the backbone models. A detailed derivation of the theoretical complexities can be found in \textbf{Appendix~\ref{sec:TCD}}, and the comprehensive empirical runtime statistics are provided in \textbf{Appendix~\ref{app:runtime}}.

\subsection{Ethical Considerations}
We acknowledge the importance of ethical data usage in recommender systems. This work utilizes public, anonymized datasets (Tmall, Jdata, Taobao) to strictly protect user privacy. Furthermore, our proposed causal debiasing framework is explicitly designed to mitigate behavioral biases, contributing to more fair and unbiased recommendation outcomes.

%% file: LaTeX/chapters/chapter_6_CONCLUSION.tex
\section{Conclusion}
In this paper, we propose MCLMR, a novel model-agnostic causal learning framework that addresses multi-behavior biases in recommendation systems. We first construct a causal graph to model the complex confounding effects from user behavioral habits and item multi-behavior distributions. Then we design an Adaptive Aggregation module based on Mixture-of-Experts to dynamically fuse auxiliary behaviors, and a Bias-aware Contrastive Learning module with user-item dual alignment to bridge semantic gaps while accounting for bias distortions. Extensive experiments on three real-world datasets demonstrate that MCLMR consistently improves various existing MBR models, validating its effectiveness.

%% file: LaTeX/chapters/chapter_7_APPENDIX.tex
\appendix

\section{Causal Effect Derivation}
\label{sec:causal_effect}

To block the backdoor paths created by confounding biases, we apply causal intervention using Pearl's do-calculus. We aim to compute the interventional probability $P(Y|do(I), do(U))$, which represents the unbiased user preference. The derivation, based on the intervened graph $G_{do}$, is as follows:

\begin{subequations}
\label{equ:P_do_IU}\footnotesize
\begin{align}
& P(Y|do(I), do(U)) \notag \\
& = P_{G_{do}}(Y|I,U) \\
& = \sum_{m,B_{i},B_{u}}P_{G_{do}}(Y,M=m,B_{i},B_{u}|I,U) \\
& = \sum_{m,B_{i},B_{u}}P_{G_{do}}(Y|M, B_{i}, B_{u}, I, U) \nonumber \\
& \qquad \times P_{G_{do}}(M|B_{i}, B_{u}, I, U) \nonumber \\
& \qquad \times P_{G_{do}}(B_{i}|I,U)P_{G_{do}}(B_{u}|B_{i},I,U) \\
& = \sum_{m,B_{i},B_{u}}P(Y|M,I,U)P(M|B_{i},B_{u},I,U) \notag\\
&\qquad \times P(B_{i})P(B_{u}) \\
& = \sum_{B_{i}}P(B_{i})\sum_{B_{u}}P(B_{u})\sum_{m}P(Y|M=m,I,U) \nonumber \\
& \qquad \times P(M=m|B_{i},B_{u},I,U)
\end{align}
\end{subequations}

The derivation steps are justified as follows:
\begin{itemize}
    \item (\ref{equ:P_do_IU}a) is based on the definition of the do-operator, which equates an intervention on a variable to conditioning on it in the manipulated graph $G_{do}$.
    \item (\ref{equ:P_do_IU}b) is due to the law of total probability, marginalizing over the mediator variables $M$ (user's real-time intention), $B_i$ (item's bias), and $B_u$ (user's bias).
    \item (\ref{equ:P_do_IU}c) applies the chain rule of probability to the joint probability distribution within the summation.
    \item (\ref{equ:P_do_IU}d) is based on the conditional independencies in $G_{do}$. The interventions $do(I)$ and $do(U)$ block the backdoor paths, making the outcome $Y$ independent of the biases $B_i, B_u$ given $\{M, I, U\}$. Since $B_i$ and $B_u$ are exogenous, they are independent of the interventions. We then apply the mechanism invariance assumption to replace probabilities from $G_{do}$ with those from the observational graph $G$.
    \item (\ref{equ:P_do_IU}e) is a rearrangement of the terms from the previous step to arrive at the final backdoor adjustment formula, which corresponds to Equation 2 in the main paper.
\end{itemize}

\section{Hyperparameter Settings}
\label{sec:hyperparameter}
We conduct a comprehensive grid search for MCLMR's specific hyperparameters. The key search spaces are:
\begin{itemize}
    \item \textbf{Causal Debiasing ($\gamma_u$, $\gamma_i$)}: The coefficients for user and item multi-behavior biases are searched within $\{0.01, 0.1, 0.5, 1.0\}$.
    \item \textbf{MoE Gate}: The hidden dimension for experts is tuned amongst $\{32, 64, 128\}$.
    \item \textbf{Contrastive Learning}: The temperature $\tau$ and weight $\lambda_{CL}$ are tuned to balance the auxiliary task.
\end{itemize}
Final configurations for reproducibility are available in the released source code.

\section{Detailed Analysis of Bias-aware Temperature}
\label{app:temperature}

This section analyzes the rationale and empirical superiority of our bias-aware temperature mechanism.

\subsection{Theoretical Justification}
Temperature $\tau$ in InfoNCE controls distribution smoothness and hard negative penalties. Our design balances supervised (BPR) and self-supervised (CL) tasks: (1) Redundancy in High-Bias Behaviors: High-bias entities (e.g., active users) dominate BPR gradient updates. To reduce redundancy in the auxiliary task, we assign them a lower temperature to sharpen the distribution. (2) Signal in Low-Bias Behaviors: Sparse behaviors provide weak signals. We assign them a higher temperature (closer to $\tau_0$) to maintain a softer distribution, preventing overfitting to noise offering auxiliary supervision.

\subsection{Experimental Variants}
We compared our strategy (Eq~\ref{tau} : $\tau = \tau_0 \cdot (1 - \alpha \cdot b)$) against four variants: (1) Fixed Temperature: $\tau$ is set to a constant scalar (e.g., 0.1). (2) Learnable Temperature: $\tau$ is a trainable parameter optimized via backpropagation. (3) Random Temperature: $\tau$ is sampled from $U(0.1, 0.5)$, inspired by SimGCL. (4) Alternative Mappings: Linear ($\tau = \tau_0 (1 + \alpha b)$) and Inverse ($\tau = \tau_0 (1 - \alpha / b)$) mappings.

\subsection{Performance Comparison}
Table \ref{tab:temp_analysis} shows the performance on the CRGCN backbone.

\begin{table}[h]
  \caption{Performance comparison of temperature strategies on CRGCN (HR@50 and NDCG@50).}
  \label{tab:temp_analysis}
  \centering
  \resizebox{\linewidth}{!}{
  \begin{tabular}{l|cc|cc|cc}
    \toprule
    \multirow{2}{*}{\textbf{Method Variant}} & \multicolumn{2}{c|}{\textbf{Jdata}} & \multicolumn{2}{c|}{\textbf{Taobao}} & \multicolumn{2}{c}{\textbf{Tmall}} \\
    & HR@50 & NDCG@50 & HR@50 & NDCG@50 & HR@50 & NDCG@50 \\
    \midrule
    Variant 1: Fixed Temperature & 0.7150 & 0.3414 & 0.2631 & 0.1091 & 0.2650 & 0.1009 \\
    Variant 2: Alt. Mapping ($1+\alpha \cdot b$) & 0.6422 & 0.3213 & 0.2623 & 0.1098 & 0.2622 & 0.1035 \\
    Variant 3: Learnable Temperature & 0.6775 & 0.3419 & 0.2481 & 0.1008 & 0.2703 & 0.1044 \\
    Variant 4: Random Temperature & 0.6970 & 0.3171 & 0.2464 & 0.0990 & 0.2574 & 0.0957 \\
    \midrule
    \textbf{MCLMR (Ours)} & \textbf{0.7295} & \textbf{0.3801} & \textbf{0.2736} & \textbf{0.1120} & \textbf{0.2753} & \textbf{0.1075} \\
    \bottomrule
  \end{tabular}
  }
\end{table}

\subsubsection{Analysis}
Our bias-aware strategy consistently outperforms all variants (Table \ref{tab:temp_analysis}): (1) vs. Fixed \& Random: Random temperature (Var 4) lacks behavioral priors, while Fixed temperature (Var 1) ignores heterogeneity of user activity. (2) vs. Learnable: Directly learning $\tau$ (Var 3) underperforms, suggesting without explicit bias-aware guidance, the model fails to optimally balance BPR and CL tasks. (3) vs. Alternatives: The performance drop in Variant 2 confirms that lowering $\tau$ for high-bias instances is the correct direction.

These results validate that incorporating a bias-aware prior is crucial for multi-behavior contrastive learning.

\section{Detailed Analysis with Other Backbones}
\label{sec:appendix_analysis}

In this section, we provide supplementary results for our in-depth analysis, using HECGCN and LightGCN as alternative GCN backbones. This complements the primary analysis in the main text, which utilizes CRGCN. The experimental setup for user/item group division and evaluation methodology remains consistent with that described in the main paper.

\subsection{Analysis with HECGCN Backbone}

We first replace the CRGCN backbone with HECGCN to validate the generalizability of our model's improvements.

\subsubsection{Performance in User Groups}
\begin{figure}[h!]
    \centering
    \includegraphics[width=0.50\textwidth]{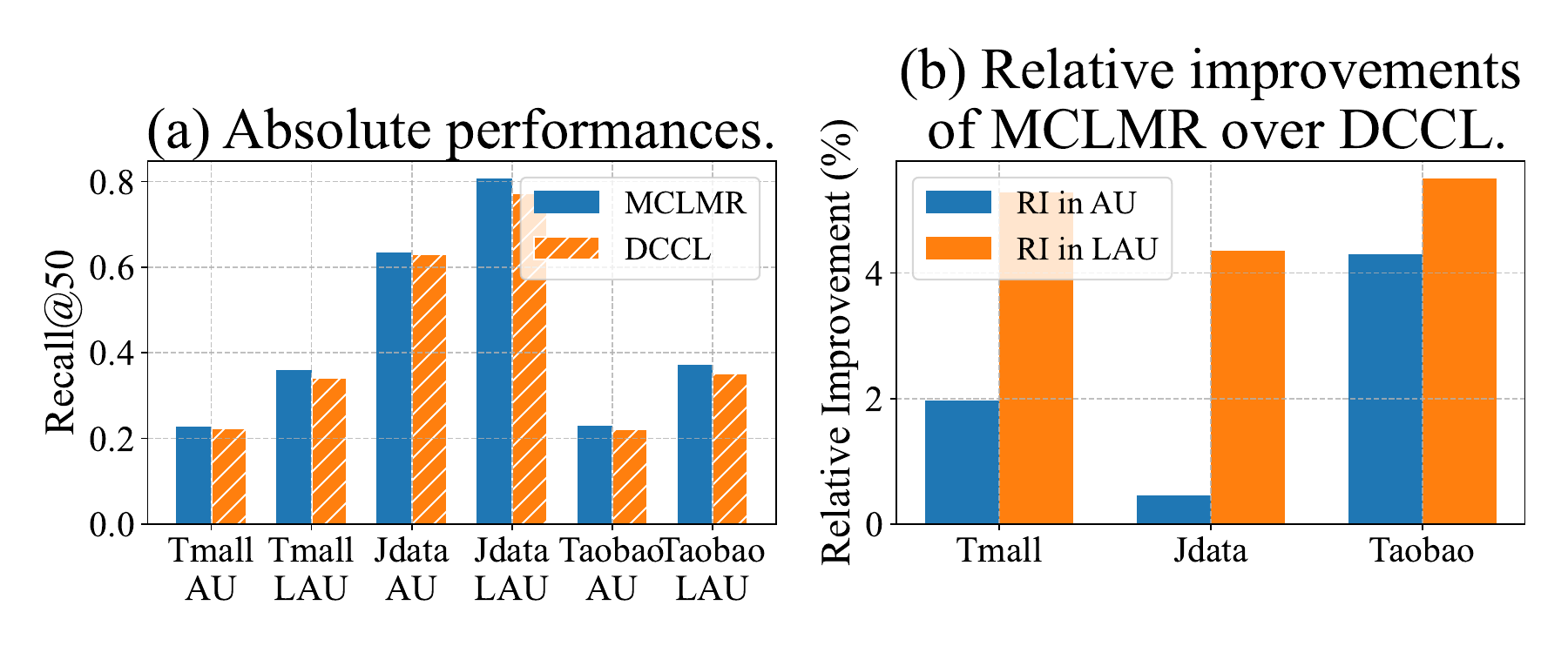} 
    \caption{Performance comparison with the \textbf{HECGCN} backbone in active (AU) and less-active (LAU) user groups. (a) Absolute performance; (b) Relative improvements of MCLMR over DCCL.}
    \label{fig:appendix_hecgcn_user}
\end{figure}

As shown in Figure~\ref{fig:appendix_hecgcn_user}, the results with the HECGCN backbone are consistent with our findings. Our proposed model (HECGCN+MCLMR) consistently outperforms the baseline (HECGCN+DCCL) in both active and less-active user groups. More importantly, the relative improvement is more pronounced in the less-active user group, once again demonstrating our model's effectiveness in alleviating data sparsity issues.

\subsubsection{Recommendation Quality}
\begin{figure}[h!]
    \centering
    \includegraphics[width=0.50\textwidth]{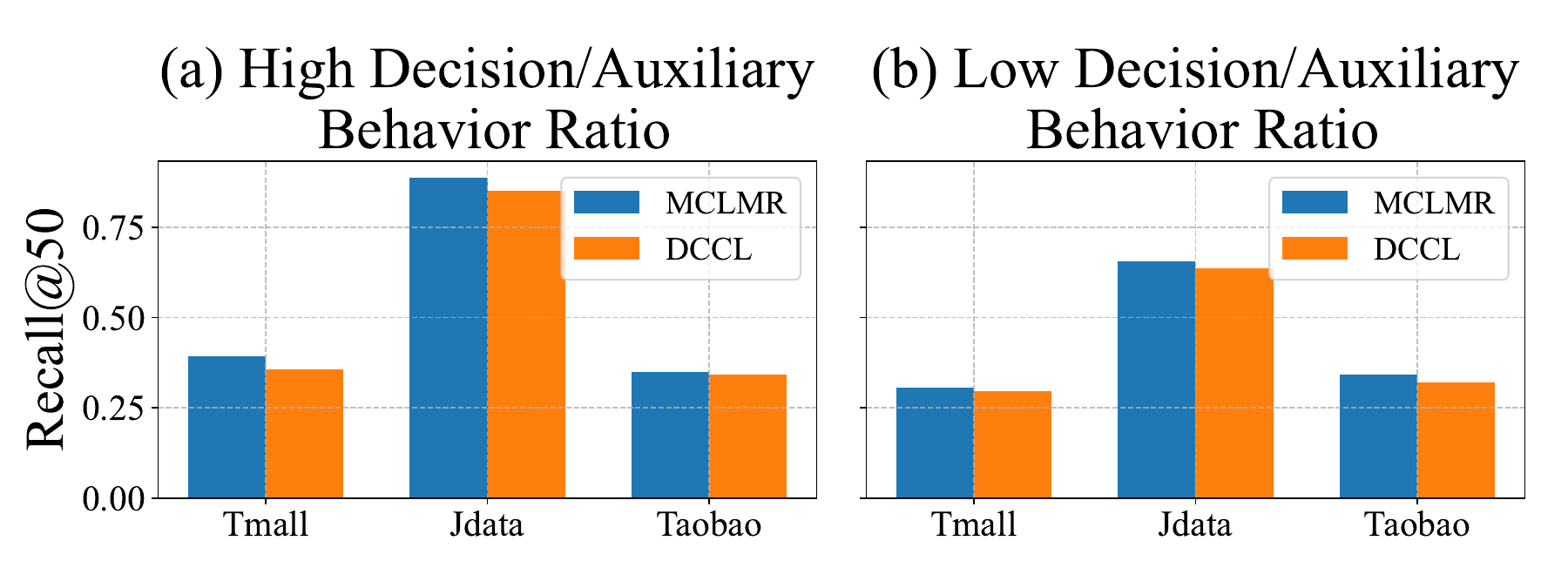} 
    \caption{Recall on high and low decision/auxiliary behavior ratio item groups with the \textbf{HECGCN} backbone.}
    \label{fig:appendix_hecgcn_item}
\end{figure}

Figure~\ref{fig:appendix_hecgcn_item} illustrates the recommendation quality on items with different decision/auxiliary behavior ratios. When built upon HECGCN, our model achieves significant gains in recommending high-quality items (high-ratio group) and maintains a strong advantage in the low-ratio group. This further validates that our method's ability to mitigate confounders is robust and not limited to a single GCN architecture.

\subsection{Analysis with LightGCN Backbone}

We further conduct experiments using LightGCN, a simpler yet powerful backbone, to verify our model's compatibility and effectiveness.

\subsubsection{Performance in User Groups}
\begin{figure}[h!]
    \centering
    \includegraphics[width=0.50\textwidth]{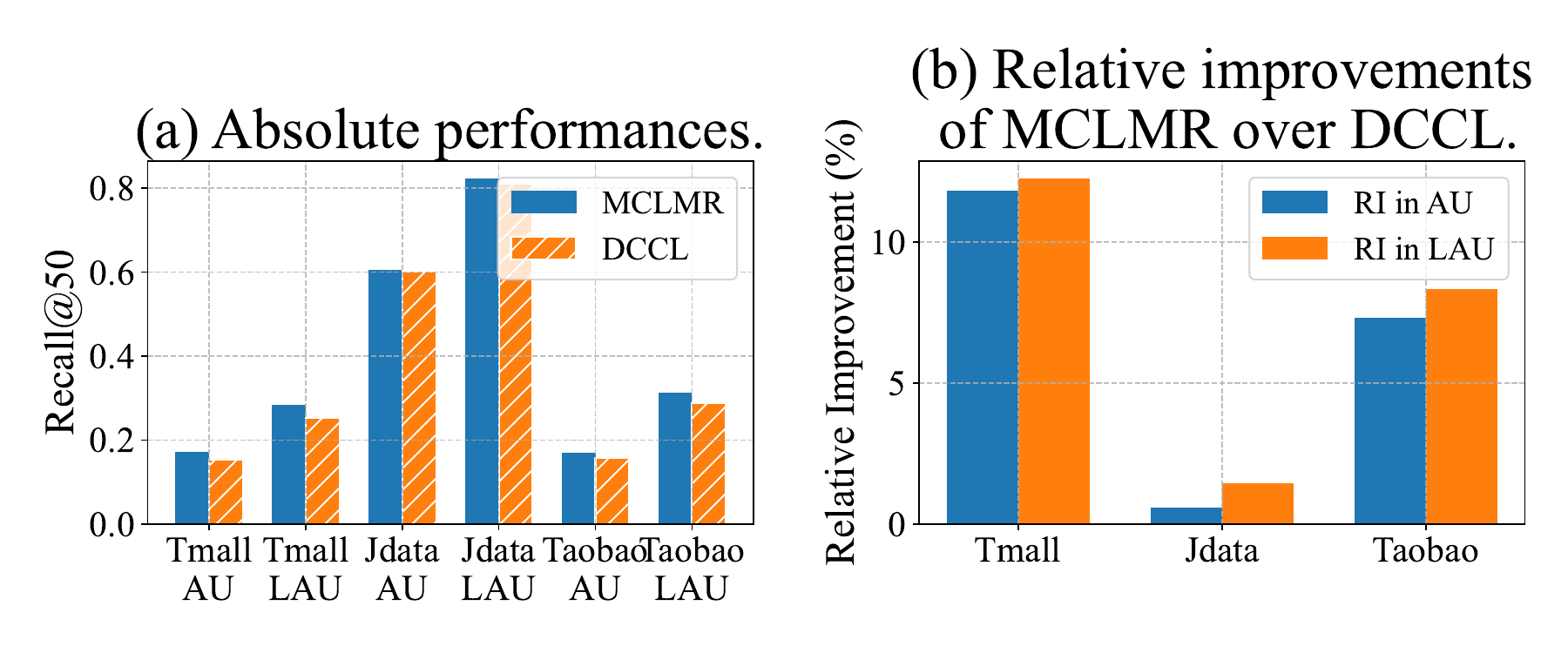} 
    \caption{Performance comparison with the \textbf{LightGCN} backbone in active (AU) and less-active (LAU) user groups. (a) Absolute performance; (b) Relative improvements of MCLMR over DCCL.}
    \label{fig:appendix_lightgcn_user}
\end{figure}

The results, summarized in Figure~\ref{fig:appendix_lightgcn_user}, show a similar pattern. The LightGCN+MCLMR combination surpasses the LightGCN+DCCL baseline across both user segments. The greater relative improvement in the less-active user group highlights our model's strong synergy with different types of graph neural networks, from complex (CRGCN, HECGCN) to more streamlined ones (LightGCN).

\subsubsection{Recommendation Quality}
\begin{figure}[h!]
    \centering
    \includegraphics[width=0.50\textwidth]{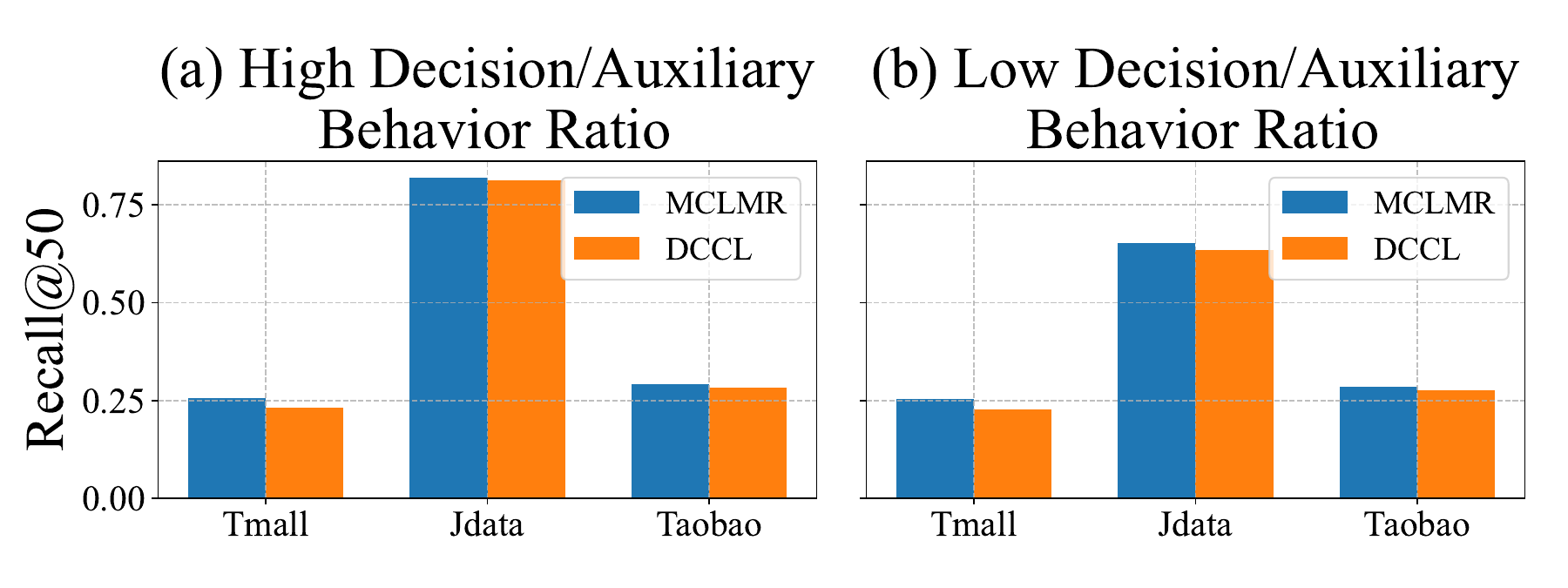} 
    \caption{Recall on high and low decision/auxiliary behavior ratio item groups with the \textbf{LightGCN} backbone.}
    \label{fig:appendix_lightgcn_item}
\end{figure}

As depicted in Figure~\ref{fig:appendix_lightgcn_item}, our model continues to excel at identifying high-quality items when paired with LightGCN. It demonstrates superior performance in the high-ratio item group while also performing robustly in the low-ratio group. This consistency across all three backbones provides strong evidence that the improvements stem from our proposed multi-level causal learning framework rather than the specifics of the GCN architecture.

\section{Time Complexity Derivation}
\label{sec:TCD}
Here, we provide a detailed derivation of the computational complexity introduced by the MCLMR framework and analyze the complexity of baseline models.

\subsection{Notations}
We define the following symbols for the complexity analysis:
\begin{itemize}
    \item $N, U, I, E_{\text{total}}$: Total number of nodes ($N=U+I$), users, items, and edges across all behavior graphs.
    \item $d, B, K, k_{exp}$: Embedding dimension, behavior types, GCN layers, and MoE experts.
    \item $u_s, b_s$: Number of unique users in a batch and the batch size.
    \item $r, L_{\text{max}}, H$: Avg. Jaccard cost, max sequence length (BCIPM), and hyperedges (HEC-GCN).
\end{itemize}

\subsection{Complexity of MCLMR Modules}
The total incremental computational overhead of our framework, $\mathcal{O}_{MCLMR}$, is composed of its three main modules as described in the paper. 

Jaccard Aggregation Module: The structural gating path uses Jaccard similarity to measure the overlap between item sets for a user across different behaviors. This calculation is performed for each user and for behaviors, leading to a complexity of $\mathcal{O}(N \cdot B^2 \cdot r)$. 

MoE Gating Aggregation Module: The semantic gating path employs a Mixture-of-Experts (MoE) network. For each user and each pair of behaviors, the MoE gate processes the embeddings. The complexity is proportional to the number of users, the square of the number of behaviors, the number of experts, and the embedding dimension, resulting in $\mathcal{O}(N \cdot B^2 \cdot k_{exp} \cdot d)$. 

Multi-behavior Contrastive Learning Module: The relational alignment is achieved via a contrastive loss calculated over pairs of different behavioral views for each user in a batch. For each of the $u_s$ users in a batch, we compute similarity scores for all $B(B-1)/2$ pairs of behaviors. This results in a complexity per batch of $\mathcal{O}(u_s \cdot B^2 \cdot d^2)$. Symmetrically, for each of the $i_s$ unique items in the batch, the item-centric complexity is $\mathcal{O}(i_s \cdot B^2 \cdot d^2)$.

Summing the complexities of the components, the total incremental complexity introduced by the MCLMR framework is:
$$ \mathcal{O}_{MCLMR} = \mathcal{O}(N \cdot B^2 \cdot (r + k_{exp} \cdot d) + (u_s + i_s) \cdot B^2 \cdot d^2) $$

\subsection{Complexity of Backbone Models}
The computational complexity of each backbone model is determined by its core architecture and key operations, as detailed below:
(1) CRGCN: The complexity is primarily determined by its graph propagation mechanism. The model employs a cascading structure with $L$ GCN layers for each of the $K$ behaviors. Since graph propagation is the most computationally intensive step, the dominant complexity is driven by it, which is $\mathcal{O}(L \cdot E_{\text{total}} \cdot d)$. (2) HEC-GCN: The model's complexity is composed of three main parts. (1) Graph Convolution: Standard GCN propagation on user-item graphs, with a complexity of $\mathcal{O}(L \cdot E_{\text{total}} \cdot d)$. (2) Hypergraph Learning: Operations on the behavior-specific hypergraphs, costing approximately $\mathcal{O}(K \cdot (M+N) \cdot S \cdot d)$. (3) Contrastive Learning: The theoretically dominant component, which compares each node against all others, resulting in a complexity of $\mathcal{O}(K \cdot (M^2 + N^2) \cdot d)$. In practice, this is often optimized using in-batch negative sampling to $\mathcal{O}(K \cdot b^2 \cdot d)$ per batch, where $b$ is the batch size. Thus, the total theoretical complexity is: $ \mathcal{O}(L \cdot E_{\text{total}} \cdot d + K \cdot (M+N) \cdot S \cdot d + K \cdot (M^2 + N^2) \cdot d) $ (3)BCIPM: The model's complexity relies on GCNs for representation learning and a specialized neighborhood aggregation mechanism. The total complexity for graph propagation is approximately $\mathcal{O}(L \cdot E_{\text{total}} \cdot d)$. The key computational module is the neighborhood aggregation, which for a batch of $b_s$ users processing padded neighbor lists up to $L_{\text{max}}$, has a complexity of $\mathcal{O}(b_s \cdot L_{\text{max}} \cdot d^2)$. Therefore, the final complexity is dominated by these two steps: $\mathcal{O}(L \cdot E_{\text{total}} \cdot d + b_s \cdot L_{\text{max}} \cdot d^2)$.

\section{Empirical Runtime Analysis}
\label{app:runtime}

To complement the theoretical complexity derivation in Appendix~\ref{sec:TCD}, we conducted empirical tests to measure the real-world computational overhead and inference latency.

\subsection{Experimental Setup}
All runtime experiments were conducted on a Linux server equipped with an NVIDIA H20-3e GPU and Intel Xeon CPUs. We measured the average time per training epoch and the total inference time for generating recommendations for all users on the test set.

\subsection{Training Overhead}
Table \ref{tab:runtime} compares the training time per epoch of the backbone models with and without MCLMR integration on the Tmall and Taobao datasets.

\begin{table}[h]
\caption{Comparison of average training time (seconds per epoch) on NVIDIA H20-3e GPU.}
\label{tab:runtime}
\centering
\begin{tabular}{l|cc|cc}
\toprule
\multirow{2}{*}{\textbf{Method}} & \multicolumn{2}{c|}{\textbf{Tmall}} & \multicolumn{2}{c}{\textbf{Taobao}} \\
& Original & +MCLMR & Original & +MCLMR \\
\midrule
HEC-GCN & 625s & 650s & 502s & 579s \\
CRGCN & 145s & 182s & 110s & 135s \\
\bottomrule
\end{tabular}
\end{table}

The results show that MCLMR introduces a manageable overhead. For complex backbones like HEC-GCN on Tmall, the increase is approximately 4\% (625s to 650s). Even for lighter backbones like CRGCN, the absolute time increase is small, validating that our framework is computationally efficient for large-scale training.

\subsection{Inference Latency}
Crucially, the additional modules in MCLMR (e.g., MoE Gating, Contrastive Learning) are designed to refine the user and item embeddings during the \textit{training phase}. 

At the \textit{inference phase}, the recommendation score is computed using the standard inner product of the learned embeddings (Eq. 13 in the main text):
$$ \hat{y}_{ui} = e_{u,K_t} \cdot e_{i,K_t}^T $$
This operation is identical to that of the backbone models (e.g., Matrix Factorization or LightGCN style retrieval). Therefore, MCLMR incurs zero additional latency during online inference, making it highly suitable for real-time recommendation scenarios where low latency is critical.


%% file: ref.bib
@String{Computing = "Computing" }

@String{Computer = "{IEEE} Computer" }

@String{Springer = "Springer-Verlag" }

@ARTICLE{2018arXiv180806581W,
       author = {{Wang}, Yixin and {Liang}, Dawen and {Charlin}, Laurent and {Blei}, David M.},
        title = "{The Deconfounded Recommender: A Causal Inference Approach to Recommendation}",
      journal = {arXiv e-prints},
         year = 2018,
        month = aug,
          eid = {arXiv:1808.06581},
        pages = {arXiv:1808.06581},
          doi = {10.48550/arXiv.1808.06581},
archivePrefix = {arXiv},
       eprint = {1808.06581},
 primaryClass = {cs.IR},
       adsurl = {https://ui.adsabs.harvard.edu/abs/2018arXiv180806581W}
}

@article{10.1016/j.ins.2024.120834,
author = {Liao, Jie and Yang, Min and Zhou, Wei and Zhang, Hongyu and Wen, Junhao},
title = {Modeling item exposure and user satisfaction for debiased recommendation with causal inference},
year = {2024},
issue_date = {Aug 2024},
publisher = {Elsevier Science Inc.},
address = {USA},
volume = {676},
number = {C},
issn = {0020-0255},
url = {https://doi.org/10.1016/j.ins.2024.120834},
doi = {10.1016/j.ins.2024.120834},
journal = {Inf. Sci.},
month = aug,
numpages = {15}
}

@misc{krauth2022breakingfeedbackloopsrecommender,
      title={Breaking Feedback Loops in Recommender Systems with Causal Inference}, 
      author={Karl Krauth and Yixin Wang and Michael I. Jordan},
      year={2022},
      eprint={2207.01616},
      archivePrefix={arXiv},
      primaryClass={cs.IR},
      url={https://arxiv.org/abs/2207.01616}, 
}

@article{10.1145/3564284,
author = {Chen, Jiawei and Dong, Hande and Wang, Xiang and Feng, Fuli and Wang, Meng and He, Xiangnan},
title = {Bias and Debias in Recommender System: A Survey and Future Directions},
year = {2023},
issue_date = {July 2023},
publisher = {Association for Computing Machinery},
address = {New York, NY, USA},
volume = {41},
number = {3},
issn = {1046-8188},
url = {https://doi.org/10.1145/3564284},
doi = {10.1145/3564284},
journal = {ACM Trans. Inf. Syst.},
month = feb,
articleno = {67},
numpages = {39}
}

@inproceedings{10.1145/3627673.3679763,
author = {Mansoury, Masoud and Mobasher, Bamshad and van Hoof, Herke},
title = {Mitigating Exposure Bias in Online Learning to Rank Recommendation: A Novel Reward Model for Cascading Bandits},
year = {2024},
isbn = {9798400704369},
publisher = {Association for Computing Machinery},
address = {New York, NY, USA},
url = {https://doi.org/10.1145/3627673.3679763},
doi = {10.1145/3627673.3679763},
booktitle = {Proceedings of the 33rd ACM International Conference on Information and Knowledge Management},
pages = {1638–1648},
numpages = {11},
location = {Boise, ID, USA},
series = {CIKM '24}
}

@ARTICLE{2020arXiv200615772A,
       author = {{Abdollahpouri}, Himan and {Mansoury}, Masoud},
        title = "{Multi-sided Exposure Bias in Recommendation}",
      journal = {arXiv e-prints},
         year = 2020,
        month = jun,
          eid = {arXiv:2006.15772},
        pages = {arXiv:2006.15772},
          doi = {10.48550/arXiv.2006.15772},
archivePrefix = {arXiv},
       eprint = {2006.15772},
 primaryClass = {cs.IR},
       adsurl = {https://ui.adsabs.harvard.edu/abs/2020arXiv200615772A}
}

@inproceedings{NIPS2015_39027dfa,
 author = {Swaminathan, Adith and Joachims, Thorsten},
 booktitle = {Advances in Neural Information Processing Systems},
 editor = {C. Cortes and N. Lawrence and D. Lee and M. Sugiyama and R. Garnett},
 pages = {},
 publisher = {Curran Associates, Inc.},
 title = {The Self-Normalized Estimator for Counterfactual Learning},
 url = {https://proceedings.neurips.cc/paper_files/paper/2015/file/39027dfad5138c9ca0c474d71db915c3-Paper.pdf},
 volume = {28},
 year = {2015}
}

@inproceedings{10.1145/3543873.3584637,
author = {Zhao, Weiqi and Tang, Dian and Chen, Xin and Lv, Dawei and Ou, Daoli and Li, Biao and Jiang, Peng and Gai, Kun},
title = {Disentangled Causal Embedding With Contrastive Learning For Recommender System},
year = {2023},
isbn = {9781450394192},
publisher = {Association for Computing Machinery},
address = {New York, NY, USA},
url = {https://doi.org/10.1145/3543873.3584637},
doi = {10.1145/3543873.3584637},
booktitle = {Companion Proceedings of the ACM Web Conference 2023},
pages = {406–410},
numpages = {5},
location = {Austin, TX, USA},
series = {WWW '23 Companion}
}

@inproceedings{10.1145/3240323.3240360,
author = {Bonner, Stephen and Vasile, Flavian},
title = {Causal embeddings for recommendation},
year = {2018},
isbn = {9781450359016},
publisher = {Association for Computing Machinery},
address = {New York, NY, USA},
url = {https://doi.org/10.1145/3240323.3240360},
doi = {10.1145/3240323.3240360},
booktitle = {Proceedings of the 12th ACM Conference on Recommender Systems},
pages = {104–112},
numpages = {9},
location = {Vancouver, British Columbia, Canada},
series = {RecSys '18}
}

@article{luo2024ci4rs,
  title = {A survey on causal inference for recommendation},
  journal = {The Innovation},
  volume = {5},
  number = {2},
  pages = {100590},
  year = {2024},
  issn = {2666-6758},
  doi = {https://doi.org/10.1016/j.xinn.2024.100590},
  url = {https://www.cell.com/the-innovation/fulltext/S2666-6758(24)00028-6},
  author = {Luo, Huishi and Zhuang, Fuzhen and Xie, Ruobing and Zhu, Hengshu and Wang, Deqing and An, Zhulin and Xu, Yongjun}
}

@article{10.1145/3745023,
author = {Chen, Yuzhe and Cao, Jie and Wang, Youquan and Wu, Jia and Chen, Huanhuan and Xu, Guandong},
title = {Causal Variational Inference for Deconfounded Multi-Behavior Recommendation},
year = {2025},
issue_date = {November 2025},
publisher = {Association for Computing Machinery},
address = {New York, NY, USA},
volume = {43},
number = {6},
issn = {1046-8188},
url = {https://doi.org/10.1145/3745023},
doi = {10.1145/3745023},
journal = {ACM Trans. Inf. Syst.},
month = sep,
articleno = {151},
numpages = {26}
}

@article{cmsr,
title = {Causal cascading convolution networks for multi-behavior sequential recommendation},
journal = {Information Sciences},
volume = {720},
pages = {122484},
year = {2025},
issn = {0020-0255},
doi = {https://doi.org/10.1016/j.ins.2025.122484},
url = {https://www.sciencedirect.com/science/article/pii/S0020025525006164},
author = {Dan Lu and Shiqing Wu and Hao Zhang and Guandong Xu and Qilong Han}
}

@article{cfsurvey,
  title={A survey of collaborative filtering techniques},
  author={Su, Xiaoyuan and Khoshgoftaar, Taghi M},
  journal={Advances in artificial intelligence},
  volume={2009},
  year={2009},
  publisher={Hindawi}
}

@inproceedings{ncf,
  title={Neural collaborative filtering},
  author={He, Xiangnan and Liao, Lizi and Zhang, Hanwang and Nie, Liqiang and Hu, Xia and Chua, Tat-Seng},
  booktitle={WWW},
  pages={173--182},
  year={2017}
}

@article{huang2021recent,
  title={Recent Advances in Heterogeneous Relation Learning for Recommendation},
  author={Huang, Chao},
  journal={arXiv preprint arXiv:2110.03455},
  year={2021}
}

@inproceedings{nmtr,
  title={Neural multi-task recommendation from multi-behavior data},
  author={Gao, Chen and He, Xiangnan and Gan, Dahua and Chen, Xiangning and Feng, Fuli and Li, Yong and Chua, Tat-Seng and Jin, Depeng},
  booktitle={ICDE},
  pages={1554--1557},
  year={2019},
  organization={IEEE}
}

@inproceedings{matn,
  title={Multiplex behavioral relation learning for recommendation via memory augmented transformer network},
  author={Xia, Lianghao and Huang, Chao and Xu, Yong and Dai, Peng and Zhang, Bo and Bo, Liefeng},
  booktitle={SIGIR},
  pages={2397--2406},
  year={2020}
}

@inproceedings{mbgcn,
  title={Multi-behavior recommendation with graph convolutional networks},
  author={Jin, Bowen and Gao, Chen and He, Xiangnan and Jin, Depeng and Li, Yong},
  booktitle={SIGIR},
  year={2020}
}

@inproceedings{cmf,
  title={Improving user topic interest profiles by behavior factorization},
  author={Zhao, Zhe and Cheng, Zhiyuan and Hong, Lichan and Chi, Ed H},
  booktitle={Proceedings of the 24th International Conference on World Wide Web},
  pages={1406--1416},
  year={2015}
}

@article{crgcn,
author = {Yan, Mingshi and Cheng, Zhiyong and Gao, Chen and Sun, Jing and Liu, Fan and Sun, Fuming and Li, Haojie},
title = {Cascading Residual Graph Convolutional Network for Multi-Behavior Recommendation},
year = {2023},
issue_date = {January 2024},
publisher = {Association for Computing Machinery},
address = {New York, NY, USA},
volume = {42},
number = {1},
issn = {1046-8188},
url = {https://doi.org/10.1145/3587693},
doi = {10.1145/3587693},
month = aug,
articleno = {10},
numpages = {26}
}

@article{bpr,
  title={BPR: Bayesian personalized ranking from implicit feedback},
  author={Rendle, Steffen and Freudenthaler, Christoph and Gantner, Zeno and Schmidt-Thieme, Lars},
  journal={arXiv preprint arXiv:1205.2618},
  year={2012}
}

@inproceedings{rgcn,
  title={Modeling relational data with graph convolutional networks},
  author={Schlichtkrull, Michael and Kipf, Thomas N and Bloem, Peter and Van Den Berg, Rianne and Titov, Ivan and Welling, Max},
  booktitle={European semantic web conference},
  pages={593--607},
  year={2018},
  organization={Springer}
}

@article{CounterfactualReasoningBottou2013,
author = {Bottou, L\'{e}on and Peters, Jonas and Qui\~{n}onero-Candela, Joaquin and Charles, Denis X. and Chickering, D. Max and Portugaly, Elon and Ray, Dipankar and Simard, Patrice and Snelson, Ed},
title = {Counterfactual reasoning and learning systems: the example of computational advertising},
year = {2013},
issue_date = {January 2013},
publisher = {JMLR.org},
volume = {14},
number = {1},
issn = {1532-4435},
journal = {J. Mach. Learn. Res.},
month = jan,
pages = {3207–3260},
numpages = {54}
}

@misc{chen2021biasdebiasrecommendersystem,
      title={Bias and Debias in Recommender System: A Survey and Future Directions}, 
      author={Jiawei Chen and Hande Dong and Xiang Wang and Fuli Feng and Meng Wang and Xiangnan He},
      year={2021},
      eprint={2010.03240},
      archivePrefix={arXiv},
      primaryClass={cs.IR},
      url={https://arxiv.org/abs/2010.03240}, 
}

@inproceedings{gnmr,
  title={Multi-Behavior Enhanced Recommendation with Cross-Interaction Collaborative Relation Modeling},
  author={Xia, Lianghao and Huang, Chao and Xu, Yong and Dai, Peng and Lu, Mengyin and Bo, Liefeng},
  booktitle={ICDE},
  pages={1931--1936},
  year={2021},
  organization={IEEE}
}

@article{mbhgcn,
  title={MB-HGCN: A hierarchical graph convolutional network for multi-behavior recommendation},
  author={Yan, Mingshi and Cheng, Zhiyong and Sun, Jing and Sun, Fuming and Peng, Yuxin},
  journal={arXiv preprint arXiv:2306.10679},
  year={2023}
}

@inproceedings{smbrec,
  title     = {Self-supervised Graph Neural Networks for Multi-behavior Recommendation},
  author    = {Gu, Shuyun and Wang, Xiao and Shi, Chuan and Xiao, Ding},
  booktitle = {Proceedings of the Thirty-First International Joint Conference on
               Artificial Intelligence, {IJCAI-22}},
  pages     = {2052--2058},
  year      = {2022},
  month     = {7},
}

@article{cai2025neighborhood,
title = {Neighborhood structure enhancement and denoising method for multi-behavior recommendation},
journal = {Neural Networks},
volume = {191},
pages = {107760},
year = {2025},
issn = {0893-6080},
doi = {https://doi.org/10.1016/j.neunet.2025.107760},
url = {https://www.sciencedirect.com/science/article/pii/S0893608025006409},
author = {Wei Cai and ZhiHong Zheng and Xuan Zhang and Weiyi Shang and Yubin Ma and WenJie Gao and Zhi Jin}
}

@article{adam,
  title={Adam: A method for stochastic optimization},
  author={Kingma, Diederik P and Ba, Jimmy},
  journal={arXiv preprint arXiv:1412.6980},
  year={2014}
}

@inproceedings{kmclr, 
author = {Xuan, Hongrui and Liu, Yi and Li, Bohan and Yin, Hongzhi}, 
title = {Knowledge Enhancement for Contrastive Multi-Behavior Recommendation}, 
year = {2023}, 
booktitle = {Proceedings of the Sixteenth ACM International Conference on Web Search and Data Mining}, 
pages = {195–203}
}

@article{lightgcn,
  title={LightGCN: Simplifying and Powering Graph Convolution Network for Recommendation},
  author={He, Xiangnan and Deng, Kuan and Wang, Xiang and Li, Yan and Zhang, Yongdong and Wang, Meng},
  journal={arXiv preprint arXiv:2002.02126},
  year={2020}
}

@inproceedings{bipn, 
author = {Yan, Mingshi and Liu, Fan and Sun, Jing and Sun, Fuming and Cheng, Zhiyong and Han, Yahong}, 
title = {Behavior-Contextualized Item Preference Modeling for Multi-Behavior Recommendation}, 
booktitle = {Proceedings of the 47th International ACM SIGIR Conference on Research and Development in Information Retrieval},
year = {2024},  
pages = {946–955}, 
}

@misc{yin2024hecgcnhypergraphenhancedcascading,
      title={HEC-GCN: Hypergraph Enhanced Cascading Graph Convolution Network for Multi-Behavior Recommendation}, 
      author={Yabo Yin and Xiaofei Zhu and Wenshan Wang and Yihao Zhang and Pengfei Wang and Yixing Fan and Jiafeng Guo},
      year={2024},
      eprint={2412.14476},
      archivePrefix={arXiv},
      primaryClass={cs.IR},
      url={https://arxiv.org/abs/2412.14476}, 
}

@article{DBLP:journals/corr/abs-1910-09337,
  author       = {Wenhao Zhang and
                  Wentian Bao and
                  Xiao{-}Yang Liu and
                  Keping Yang and
                  Quan Lin and
                  Hong Wen and
                  Ramin Ramezani},
  title        = {A Causal Perspective to Unbiased Conversion Rate Estimation on Data
                  Missing Not at Random},
  journal      = {CoRR},
  volume       = {abs/1910.09337},
  year         = {2019},
  url          = {http://arxiv.org/abs/1910.09337},
  eprinttype    = {arXiv},
  eprint       = {1910.09337},
  bibsource    = {dblp computer science bibliography, https://dblp.org}
}

@inproceedings{10.1145/3404835.3462875,
author = {Zhang, Yang and Feng, Fuli and He, Xiangnan and Wei, Tianxin and Song, Chonggang and Ling, Guohui and Zhang, Yongdong},
title = {Causal Intervention for Leveraging Popularity Bias in Recommendation},
year = {2021},
isbn = {9781450380379},
publisher = {Association for Computing Machinery},
address = {New York, NY, USA},
url = {https://doi.org/10.1145/3404835.3462875},
doi = {10.1145/3404835.3462875},
booktitle = {Proceedings of the 44th International ACM SIGIR Conference on Research and Development in Information Retrieval},
pages = {11–20},
numpages = {10},
location = {Virtual Event, Canada},
series = {SIGIR '21}
}

@inproceedings{wei2021model,
  title={Model-Agnostic Counterfactual Reasoning for Eliminating Popularity Bias in Recommender System},
  author={Wei, Tianxin and Feng, Fuli and Chen, Jiawei and Wu, Ziwei and Yi, Jinfeng and He, Xiangnan},
  booktitle={Proceedings of the 27th ACM SIGKDD Conference on Knowledge Discovery \& Data Mining},
  pages={1791--1800},
  year={2021}
}

@book{10.5555/1642718,
author = {Pearl, Judea},
title = {Causality: Models, Reasoning and Inference},
year = {2009},
isbn = {052189560X},
publisher = {Cambridge University Press},
address = {USA},
edition = {2nd}
}

@ArtifactSoftware{R,
    title = {R: A Language and Environment for Statistical Computing},
    author = {{R Core Team}},
    organization = {R Foundation for Statistical Computing},
    address = {Vienna, Austria},
    year = {2019},
    url = {https://www.R-project.org/},
}

@inproceedings{DBLP:conf/iclr/XuCLL0Y24,
  author       = {Ziqi Xu and
                  Debo Cheng and
                  Jiuyong Li and
                  Jixue Liu and
                  Lin Liu and
                  Kui Yu},
  title        = {Causal Inference with Conditional Front-Door Adjustment and Identifiable
                  Variational Autoencoder},
  booktitle    = {The Twelfth International Conference on Learning Representations,
                  {ICLR}},
  year         = {2024},
  url          = {https://openreview.net/forum?id=wFf9m4v7oC}
}

@inproceedings{XuCLLLW23,
  author       = {Ziqi Xu and
                  Debo Cheng and
                  Jiuyong Li and
                  Jixue Liu and
                  Lin Liu and
                  Ke Wang},
  title        = {Disentangled Representation for Causal Mediation Analysis},
  booktitle    = {Thirty-Seventh {AAAI} Conference on Artificial Intelligence, {AAAI}},
  pages        = {10666--10674},
  year         = {2023},
  url          = {https://doi.org/10.1609/aaai.v37i9.26266}
}

@inproceedings{Ma0RHC25,
  author       = {Chenglong Ma and
                  Ziqi Xu and
                  Yongli Ren and
                  Danula Hettiachchi and
                  Jeffrey Chan},
  title        = {{PUB:} An LLM-Enhanced Personality-Driven User Behaviour Simulator
                  for Recommender System Evaluation},
  booktitle    = {Proceedings of the 48th International {ACM} {SIGIR} Conference on
                  Research and Development in Information Retrieval, {SIGIR}},
  pages        = {2690--2694},
  year         = {2025},
  url          = {https://doi.org/10.1145/3726302.3730238}
}

@inproceedings{WanyanHM0C25,
  author       = {Xinye Wanyan and
                  Danula Hettiachchi and
                  Chenglong Ma and
                  Ziqi Xu and
                  Jeffrey Chan},
  title        = {Temporal-Aware User Behaviour Simulation with Large Language Models
                  for Recommender Systems},
  booktitle    = {Proceedings of the 34th {ACM} International Conference on Information
                  and Knowledge Management, {CIKM}},
  pages        = {5335--5339},
  year         = {2025},
  url          = {https://doi.org/10.1145/3746252.3760878}
}

@article{ZhangYCLLXZ25,
  author       = {Guixian Zhang and
                  Guan Yuan and
                  Debo Cheng and
                  Lin Liu and
                  Jiuyong Li and
                  Ziqi Xu and
                  Shichao Zhang},
  title        = {Deconfounding representation learning for mitigating latent confounding
                  effects in recommendation},
  journal      = {Knowledge and Information Systems},
  volume       = {67},
  number       = {7},
  pages        = {5999--6020},
  year         = {2025},
  url          = {https://doi.org/10.1007/s10115-025-02404-7}
}

@inproceedings{wang2018confidence,
  title={Confidence-aware matrix factorization for recommender systems},
  author={Wang, Chao and Liu, Qi and Wu, Runze and Chen, Enhong and Liu, Chuanren and Huang, Xunpeng and Huang, Zhenya},
  booktitle={Proceedings of the AAAI Conference on artificial intelligence},
  volume={32},
  number={1},
  year={2018}
}

@inproceedings{wang2020setrank,
  title={Setrank: A setwise bayesian approach for collaborative ranking from implicit feedback},
  author={Wang, Chao and Zhu, Hengshu and Zhu, Chen and Qin, Chuan and Xiong, Hui},
  booktitle={Proceedings of the aaai conference on artificial intelligence},
  volume={34},
  number={04},
  pages={6127--6136},
  year={2020}
}

@inproceedings{wang2021variable,
  title={Variable interval time sequence modeling for career trajectory prediction: Deep collaborative perspective},
  author={Wang, Chao and Zhu, Hengshu and Hao, Qiming and Xiao, Keli and Xiong, Hui},
  booktitle={Proceedings of the Web Conference 2021},
  pages={612--623},
  year={2021}
}

@article{wang2021personalized,
  title={Personalized and explainable employee training course recommendations: A bayesian variational approach},
  author={Wang, Chao and Zhu, Hengshu and Wang, Peng and Zhu, Chen and Zhang, Xi and Chen, Enhong and Xiong, Hui},
  journal={ACM Transactions on Information Systems (TOIS)},
  volume={40},
  number={4},
  pages={1--32},
  year={2021},
  publisher={ACM New York, NY}
}

@inproceedings{zhu2024graph,
  title={Graph signal diffusion model for collaborative filtering},
  author={Zhu, Yunqin and Wang, Chao and Zhang, Qi and Xiong, Hui},
  booktitle={Proceedings of the 47th International ACM SIGIR Conference on Research and Development in Information Retrieval},
  pages={1380--1390},
  year={2024}
}

@article{liu2025rethinking,
  title={Rethinking Popularity Bias in Collaborative Filtering via Analytical Vector Decomposition},
  author={Liu, Lingfeng and Song, Yixin and Shen, Dazhong and Yin, Bing and Li, Hao and Zhang, Yanyong and Wang, Chao},
  journal={arXiv preprint arXiv:2512.10688},
  year={2025}
}

@article{wang2025face,
  title={FACE: A General Framework for Mapping Collaborative Filtering Embeddings into LLM Tokens},
  author={Wang, Chao and Song, Yixin and Ye, Jinhui and Qin, Chuan and Shen, Dazhong and Liu, Lingfeng and Wang, Xiang and Zhang, Yanyong},
  journal={arXiv preprint arXiv:2510.15729},
  year={2025}
}
